\documentclass[ longbibliography,groupedaddress,showpacs,showkeys,amssymb,eqsecnum,aps,nofootinbib]{revtex4-2}
\usepackage[]{graphicx}
\usepackage[]{graphics}
\usepackage{amsmath}
\usepackage{epsf}

\usepackage{color}                   
\usepackage{verbatim}

\usepackage{hyperref}

\newcommand{\be}{\begin{equation}}

\newcommand{\ee}{\end{equation}}

\begin{document}

\author{F. T. Brandt}  
\email{fbrandt@usp.br}
\affiliation{Instituto de F\'{\i}sica, Universidade de S\~ao Paulo, S\~ao Paulo, SP 05508-090, Brazil}

\author{J. Frenkel}
\email{jfrenkel@if.usp.br}
\affiliation{Instituto de F\'{\i}sica, Universidade de S\~ao Paulo, S\~ao Paulo, SP 05508-090, Brazil}

\author{S. Martins-Filho}   
\email{sergiomartinsfilho@usp.br}
\affiliation{Instituto de F\'{\i}sica, Universidade de S\~ao Paulo, S\~ao Paulo, SP 05508-090, Brazil}

\author{D. G. C. McKeon}
\email{dgmckeo2@uwo.ca}
\affiliation{
Department of Applied Mathematics, The University of Western Ontario, London, Ontario N6A 5B7, Canada}
\affiliation{Department of Mathematics and Computer Science, Algoma University,
Sault Ste.~Marie, Ontario P6A 2G4, Canada}

\title{Restricting loop expansions in gauge theories coupled to matter}

\date{\today}

\begin{abstract}
    Quantizing any model in which a Lagrange multiplier (LM) field is used to restrict field configurations to those that satisfy the classical equations of motion, leads to at most one-loop radiative corrections. This approach can be used with both the Yang-Mills (YM) and Einstein-Hilbert (EH) action; the resulting theory is both renormalizable and unitary, has a positive energy spectrum and has no negative norm states contributing to physical processes. Although this approach cannot be consistently used with scalar fields alone, scalar fields can be coupled to gauge fields so that loop effects in the gauge sector are restricted to one-loop order in a way that satisfies the usual criterion for a consistent quantum field theory. The tree-level diagrams are those of the classical theory in which the metric couples to the energy-momentum tensor.

\end{abstract}

\pacs{11.15.-q}
\keywords{gauge theory; perturbation theory; renormalization}

\maketitle

\section{Introduction}

Normally the perturbative calculation of radiative effects in quantum field theory involves the so-called ``loop expansion'' with terms with no loops (the ``tree diagrams'') being the classical limit.  Some models become inconsistent at one-loop order, but others (the EH-action for gravity \cite{1} and massive YM theory \cite{2,3}) exhibit problems beginning at two-loop order.  It has been shown that by using a LM field to restrict paths being considered in the path integral to those satisfying the classical equation of motion, then the usual one-loop perturbative results are doubled and all higher loop effects are eliminated.  We are thus motivated to study in more detail how the use of a LM field affects the properties of quantized scalar field, a quantized YM field and a quantized metric when using the EH action.

For an $O(N)$ scalar model we will find that upon introducing a LM to eliminate diagrams with more than one-loop, divergences that arise at one-loop order can be absorbed by renormalizing the parameters (mass and coupling) that characterize the tree level action, but leave the parameters appearing at one-loop order unrenormalized.  This renormalization procedure does not lead to results consistent with unitarity, so the use of the LM field for scalar ``matter'' fields is not appropriate.

However, for YM theory, this problem does not arise for renormalizing the theory. As a result, the renormalization constants and hence renormalization group (RG) functions can be computed exactly for both the coupling constant and wave function renormalization \cite{4}.  Indeed, since the LM field eliminates diagrams beyond one-loop order, it is even possible to use a LM field to render a YM gauge model supplemented by a Proca mass term renormalizable, if not unitary \cite{5}.  It is of course well known that massless YM theory and massive YM theory when the mass is generated by the Higgs mechanism are both renormalizable and unitary to all orders in the loop expansion \cite{6}.  There is consequently no motivation for employing a LM field in conjunction with the YM field. However, using the LM field in conjunction with the YM action provides an illustration of how a LM field can be used with the EH action.

The EH action for General Relativity, at one-loop order, provided it is not interacting, is renormalizable when the equations of motion are satisfied \cite{1}. Beyond one-loop order \cite{7} or in interaction with scalar \cite{1}, vector \cite{8} or spinor \cite{9} fields, this is not the case.  Supergravity \cite{10}, higher derivative models \cite{11} , string theory \cite{12} and non-perturbative properties of renormalization group functions 
\cite{13} all have been invoked to resolve this issue.  Here we will demonstrate that using a LM field to restrict configurations of the metric to solutions of the classical field equations without matter fields can be used in a straightforward way to obtain a renormalizable, unitary model of the metric even when it interacts with ``matter'' fields.  Divergences coming from ``graviton'' loops are eliminated in a somewhat unusual way as they are absorbed into the LM field. 
It is not necessary for the background field to satisfy the equation of motion. 

In an introductory section we discuss some general features of using a LM field in conjunction with the path integral. The next three sections outline how a LM field can be used in scalar, YM and gravitational models respectively. We also show in the Appendix \ref{appA} how the first class constraints, present in a YM action supplemented by a LM field, can be used to derive the gauge invariances in this model.
In the  Appendix \ref{appB} we argue that, in consequence of the nilpotency
of the BRST transformations, the Lagrange multiplier gauge theories
are consistent with unitarity,
as the contribution of ghost states cancel against those of unphysical
polarizations of gauge fields. We also show that when there is a LM
field, the energy spectrum is bounded below and negative norm states
do not contribute to physical processes.

We are using the Feynman path integral to quantize both the fields
that normally appear in the classical action as well as the LM field
used to impose the classical equations of motion; in no sense though
is what we are considering a classical theory. There is no
partitioning of degrees of freedom into those that are ``classical''
and those that are ``quantum'', such as is done in the 
Koopman–von Neumann–Sudarshan approach as discussed in
Refs. \cite{44,45}. If we were to have such a partitioning, then when
employing canonical quantization, these coordinates and their
conjugate momenta corresponding to classical degrees of freedom would
commute. This is not the case in our approach; all degrees of freedom
have conventional quantum commutation relations.


\section{General Formalism for Quantization}

In this section, some general features of how a Lagrange multiplier field can be used to eliminate higher loop contributions to Green's functions are presented. This is, in most part, a review of the material in refs.
\cite{4,21}.

In general, a Lagrangian $\mathcal{L}(\phi_i)$ defines the dynamics of a field $\phi_i(x)$.  On occasion it is advantageous to introduce an auxiliary field $\Phi_j$ so that the Lagrangian for the system becomes $\mathcal{L}\left( \phi_i, \Phi_j\right)$.  The equation of motion for $\Phi_i$ leads to $\Phi_j = \Phi_j(\phi_i)$ so that
\begin{equation}\label{A.1}
\mathcal{L}(\phi_i) = \mathcal{L}\left( \phi_i, \Phi_j(\phi_i)\right).
\end{equation}

The generating functional
\begin{equation}\label{A.2}
Z[j_i] = \int D\phi_i \exp \frac{i}{\hbar} \int dx \left(\mathcal{L} (\phi_i) + j_i\phi_i\right)
\end{equation}
has a perturbative expansion that gives rise to multi-loop Feynman diagrams.  Only connected diagrams occur in the expansion of
\begin{equation}\label{A.3}
W[j_i] = -i\hbar \ln Z[j_i]
\end{equation}
while if
\begin{equation}\label{A.4}
B_i = \frac{\delta W[j_i]}{\delta j_i}
\end{equation}
then the Legendre transform of $W[j_i]$
\begin{equation}\label{A.5}
\Gamma[B_i] = W[j_i] - \int dx B_ij_i
\end{equation}
gives rise to one particle irreducible Feynman diagrams \cite{18,26}.

Together, Eqs. (\ref{A.3}-\ref{A.5}) lead to 
\begin{equation}\label{A.6a}
e^{\frac{i}{\hbar}\Gamma[B_i]} = \int D\phi_i \exp \frac{i}{k} \int dx \left( \mathcal{L}(\phi_i) + j_i (\phi_i - B_i)\right)\nonumber
\end{equation}
or, if
\begin{equation}\label{A.6}
\phi_i = B_i + Q_i
\end{equation}
\begin{equation}\label{A.7}
e^{\frac{i}{\hbar}\Gamma[B_i]} = \int D Q_i \exp \frac{i}{\hbar} \int dx \left( \mathcal{L}(B_i + Q_i) + j_i Q_i \right),
\end{equation}
where $B_i$ is a ``background field'' while $Q_i$ is a ``quantum field''.  From Eqs. (\ref{A.4}) and (\ref{A.5}), it follows that
\begin{equation}\label{A.8}
j_i = - \frac{\delta \Gamma[B_i]}{\delta B_i} .
\end{equation}
One can now make the expansions
\begin{equation}\label{A.9}
\Gamma[B_i] = \Gamma_{(0)} [B_i] + \hbar \Gamma_{(1)} [B_i] + \hbar^2 \Gamma_{(2)} [B_i] + \cdots
\end{equation}
and 
\begin{equation}\label{A.10}
\mathcal{L}\left(B_i + Q_i\right) = \mathcal{L} (B_i) + \frac{1}{1!} \mathcal{L}_{,i} (B_i) Q_i + \frac{1}{2!} \mathcal{L}_{,ij} (B_i) Q_iQ_j
+ \frac{1}{3!} \mathcal{L}_{,ijk} (B_i) Q_iQ_jQ_k + \ldots . 
\end{equation}
When Eqs. (\ref{A.8} - \ref{A.10}) are substituted into Eq. (\ref{A.7}) we find that to lowest order in $\hbar$,
\begin{equation}\label{A.11}
\Gamma_{(0)} [B_i] = \int dx \mathcal{L} (B_i) .
\end{equation}
From Eq. (\ref{A.11}), terms linear in $Q_i$ in the exponential of Eq. (\ref{A.7}) cancel.  It is not necessary to impose the equation of motion
\begin{equation}\label{A.12}
\mathcal{L}_{,i} (B_i) = 0
\end{equation}
on the background field $B_i$ to eliminate terms linear in $Q_i$.
To zeroth order in $\hbar$, we find that
\begin{equation}\label{A.13}
\Gamma_{(1)} = - i \ln\det{}^{-1/2} \mathcal{L}_{,jk} (B_i),
\end{equation}
the usual one-loop result.  Subsequent powers of $\hbar$ show that $\Gamma_{(n)}[B_i]$ $(n > 1)$ are associated with $n$-loop one-particle irreducible Feynman diagrams \cite{18,26}.  In obtaining this result, it is assumed that $\mathcal{L}(\phi_i)$ is independent of $\hbar$. This is not always the case; it has been pointed out that if $\mathcal{L}(\phi_i)$ depends on $\hbar$ then loop diagrams can possibly have non-vanishing contributions in the ``classical limit'' $\hbar \rightarrow 0$ \cite{27}.

Next we will consider the consequences of using a Lagrange multiplier (LM) field $\lambda_i$ to ensure that $\phi_i$ satisfies the equation of motion (\ref{A.12}).  The action of Eq. (\ref{A.2}) now becomes (upon setting $\hbar = 1$)
\begin{equation}\label{A.14}
Z[j_i,k_i] = \int D\phi_i D\lambda_i \exp i \int dx \left( \mathcal{L}(\phi_i) + \lambda_k \frac{\partial \mathcal{L}(\phi_i)}{\partial\phi_k} + j_i\phi_i + k_i\lambda_i\right).
\end{equation}
Since both $\phi_i$ and $\lambda_i$ are integrated in
Eq. \eqref{A.14}, this model is not semi-classical; it is fully
quantized.

The functional integral over $\lambda_i$ in Eq. (\ref{A.14}) results in a functional $\delta$-function so that
\begin{equation}\label{A.15}
Z[j_i,k_i] = \int D\phi_i \delta\left(\frac{\partial\mathcal{L}(\phi_i)}{\partial\phi_k} + k_\kappa \right) \exp i \int dx \left( \mathcal{L} (\phi_i) + j_i \phi_i \right).
\end{equation}
The functional analogue of 
\begin{equation}\label{A.16}
    \int \mathop{dx} f(x) \delta (g(x)) = \sum_{\bar{x}_i} f(\bar{x}_i) / |g^\prime (\bar{x}_i)|
\end{equation}
reduces Eq. (\ref{A.14}) to
\begin{equation}\label{A.17}
Z[j_i, k_i]= \sum_{\bar{\phi}_i} \exp i \int dx \left( \mathcal{L}(\bar{\phi}_i) + j_i \bar{\phi}_i\right) \det{}^{-1} \left( \mathcal{L}_{,jk}(\bar{\phi}_i)\right).
\end{equation}
In Eq. (\ref{A.16}), $\bar{x}_i$ is a solution to 
\begin{equation}\label{A.18}
g(\bar{x}_i) = 0
\end{equation}
while in Eq. (\ref{A.17}) $\bar{\phi}_i(x)$ satisfies
\begin{equation}\label{A.19}
\mathcal{L}_{,k} (\bar{\phi}_i) + k_k = 0.
\end{equation}

If we define
\begin{equation}\label{A.20}
W[j_i, k_i] = - i \ln Z[j_i,k_i]
\end{equation}
and then have
\begin{equation}\label{A.21}
B_i = \frac{\delta W[j_i,k_i]}{\delta j_i}
\end{equation}
\begin{equation}\label{A.22}
\Gamma[B_i, k_i] = W[j_i, k_i] - \int dx B_i j_i
\end{equation}
we find that
\begin{equation}\label{A.23}
e^{i\Gamma[B_i,k_i]} = \int DQ_i D\lambda_i \exp i \int dx \Big( \mathcal{L} (B_i + Q_i) 
+ \lambda_k \frac{\partial}{\partial Q_k} \mathcal{L}(B_i + Q_i) + j_i Q_i + k_i \lambda_i\Big),
\end{equation}
where now
\begin{equation}\label{A.24}
j_i = -\frac{\delta \Gamma[B_i,k_i]}{\delta B_i}.
\end{equation}
In general, we do not provide a background field for $ \lambda_{i} $ since it does not appear as an external field. The integral over $\lambda_i$ in Eq. (\ref{A.23}) now results in
\begin{equation}\label{A.25}
e^{i\Gamma[B_i,k_i]} = \int DQ_i \delta\left( \frac{\partial }{\partial Q_k}  \mathcal{L} (B_i + Q_i) + k_k\right) 
\exp i \int dx \left( \mathcal{L}(B_i + Q_i) + j_i Q_i \right)
\end{equation}
so that, much like Eq. (\ref{A.17}), we have
\begin{equation}\label{A.26}
= \sum_{\bar{Q}_i} \exp i \int dx \left( \mathcal{L} \left( B_i + \bar{Q}_i\right) + j_i \bar{Q}_i\right) \det{}^{-1}
\left( \frac{\partial^2 \mathcal{L}\left(B_i + \bar{Q}_i\right)}{\partial Q_k \partial Q_\ell} \right)
\end{equation}
where $\bar{Q}_i$ satisfies
\begin{equation}\label{A.27}
\frac{\partial \mathcal{L}(B_i + \bar{Q}_i)}{\partial Q_k} + k_k = 0.
\end{equation}
In the limit $k_k = \bar{Q}_i = 0$, we see from Eqs. (\ref{A.11}) and (\ref{A.13}) that on the right side of Eq. (\ref{A.26}) we have the product of all ``tree'' diagrams (the exponential) and the square of all the usual ``one-loop diagrams'' (the functional determinant), with the field $B_i$ on external legs.  No contributions beyond one-loop arise.  This is consistent with what results from a Feynman diagram expansion of 
$Z[j_i , k_i]$ \cite{4}.

The action in some models
\begin{equation}\label{A.28}
S = \int dx \mathcal{L}(\phi_i)
\end{equation}
is invariant under a ``gauge transformation''
\begin{equation}\label{A.29}
\phi_i \rightarrow \phi_i^\prime = \phi_i + H_{ij} (\phi_k)\xi_j
\end{equation}
so that
\begin{equation}\label{A.30}
\int dx \frac{\partial\mathcal{L}(\phi_i)}{\partial\phi_k} H_{k\ell} (\phi_i) \xi_\ell = 0.
\end{equation}
If we now consider the action
\begin{equation}\label{A.31}
S_\lambda = \int dx \left( \mathcal{L} (\phi_i) + \lambda_k \frac{\partial\mathcal{L}(\phi_i)}{\partial\phi_k}\right)
\end{equation}
then by Eq. (\ref{A.30}) it immediately follows that the transformation
\begin{equation}\label{A.32}
\lambda_i \rightarrow \lambda_i^\prime = \lambda_i + H_{ij} (\phi_k)\zeta_k
\end{equation}
leaves $S_\lambda$ invariant.  Furthermore, if
\begin{equation}\label{A.33}
\int dx \left( \mathcal{L}(\phi^\prime_j) + \lambda_i^\prime \frac{\partial \mathcal{L}(\phi_j^\prime)}{\partial\phi_i^\prime}\right) 
= \int dx \left( \mathcal{L}(\phi_j) + \lambda_i \frac{\partial \mathcal{L}(\phi_j)}{\partial\phi_i}\right)
\end{equation}
where $\phi_j^\prime = \phi_i + H_{ij}(\phi_k)\xi_j$ leaves $S$ in Eq. (\ref{A.28}) invariant, then it follows that
\begin{equation}\label{A.34}
\lambda_i \rightarrow \lambda_i^\prime = \lambda_k \frac{\partial \phi_i^\prime}{\partial\phi_k} 
= \lambda_i + \lambda_k \frac{\partial H_{ij}(\phi_\ell)\xi_j}{\partial \phi_k}.
\end{equation}
Together, Eqs. (\ref{A.29}) and (\ref{A.34}) as well as Eq. (\ref{A.32}) are invariances of $S_\lambda$ in Eq. (\ref{A.31})  \cite{21,21a}.

In order for the gauge transformation of Eq. (\ref{A.29}) to close under commutation of two successive gauge transformations, we must have
\begin{eqnarray}\label{A.35}
\left( \delta_A\delta_B - \delta_B\delta_A\right) \phi_i &=& \delta_A \left(H_{ij} \xi_j^B \right) - \delta_B\left( H_{ij} \xi_j^A\right) \nonumber\\
&=& \frac{\partial H_{ij}}{\partial\phi_k} H_{k\ell} \xi_\ell^A\xi_j^B - \frac{\partial H_{ij}}{\partial \phi_k} H_{k\ell} \xi_j^A \xi_\ell^B\nonumber \\
&=& H_{im} f_{mj\ell} \xi_\ell^a \xi_j^B
\end{eqnarray}
so that
\begin{equation}\label{A.36}
\frac{\partial H_{ij}}{\partial \phi_k} H_{k\ell} - \frac{\partial H_{i\ell}}{\partial \phi_k} H_{kj} = H_{im} f_{mj\ell}.
\end{equation}
For a gauge transformation to be consistent, the Jacobi identity\\
 $\left(\left[ \delta_A, \left[ \delta_B, \delta_C\right]\right] + 
\left[ \delta_B, \left[ \delta_C, \delta_A\right]\right] + \left[ \delta_C, \left[ \delta_A, \delta_B\right]\right]\right)\phi_i = 0$ 
must be satisfied.  This implies that $f_{kab}f_{\ell ck} + f_{kca} f_{\ell bk} + f_{kbc}f_{\ell ak} = 0$, provided $f_{ijk}$ is independent of $\phi_i$. 

The Faddeev-Popov procedure \cite{17} can be used to quantize a model with the action $S_\lambda$ possessing the gauge invariances of Eqs. (\ref{A.29}, \ref{A.34}) and (\ref{A.32}).  If we want to impose the same gauge restriction on $\phi_i$ and $\lambda_i$,
\begin{equation}\label{A.37}
F_{ij} \phi_j = 0 = F_{ij} \lambda_j
\end{equation}
then we begin by inserting the constant \cite{21,21a}
\begin{equation}\label{A.38}
1 = \int D\xi_i D\zeta_i \delta \Bigg[ F_{ij} \left(
\left( \begin{array}{c}
\phi_j\\ \lambda_j\end{array} \right) +
\left( \begin{array}{cc}
0 & H_{jk}\\
H_{jk} & \lambda_\ell \frac{\partial H_{jk}}{\partial \phi_\ell} \end{array}\right)
\left( \begin{array}{c}
\zeta_k \\ \xi_k\end{array}\right) \right) 
-  \left( \begin{array}{c}
p_i \\ q_i\end{array} \right) \Bigg] 
\det \left[ F_{ij} \left( \begin{array}{cc}
0 & H_{jk}\\
H_{jk} & \lambda_\ell \frac{\partial H_{jk}}{\partial \phi_\ell} \end{array}\right)\right] 
\end{equation}
into the path integral of Eq. (\ref{A.14}), followed by insertion of the constant
\begin{equation}\label{A.39}
\int Dp_j Dq_i \exp \frac{-i}{2\alpha} \int dx \left( p_ip_i + 2p_i q_i \right).
\end{equation}
If we then perform the gauge transformations of Eqs. (\ref{A.29}, \ref{A.34}, \ref{A.32}) with $\xi_i$, $\zeta_i$ replaced by 
$(-\xi_i , -\zeta_i)$ we are then left with
\begin{eqnarray}\label{A.40}
Z[j_i , k_i] &=& \int D\phi_i D\lambda_i \det \left[ F_{ij} \left(
\begin{array}{cc}
0 & H_{jk}\\
H_{jk} & \lambda_\ell \frac{\partial H_{jk}}{\partial \phi_\ell}\end{array}\right)\right]\nonumber\\
& &\exp i \int dx \left[ \mathcal{L}(\phi_i) + \lambda_k \frac{\partial\mathcal{L}(\phi_i)}{\partial\phi_k} + j_i \phi_i + k_i \lambda_i
- \frac{1}{2\alpha} \left( F_{ij}\phi_j F_{ik}\phi_k + 2 F_{ij}\phi_j F_{ik}\lambda_k \right)\right]
\end{eqnarray}
upon dropping normalization factors $\int D\xi_i D\zeta_i$.

It is possible to impose more than one gauge condition on $\phi_i, \lambda_i$.  This is useful in spin-two models in which one wants a propagator that is both traceless and transverse \cite{28}.  If in addition to the gauge conditions of Eq. (\ref{A.37}) we wish to have
\begin{equation}\label{A.41}
G_{ij} \phi_j = 0 = G_{ij} \lambda_j
\end{equation}
then into the path integral of Eq. (\ref{A.14}) we insert not just Eqs. (\ref{A.38})  , but also a constant that is found by replacing $(\zeta_i, \xi_i) \rightarrow (\rho_i, \theta_i), (p_i, q_i) \rightarrow (r_i, s_i), F_{ij} \rightarrow G_{ij}$ in Eq. (\ref{A.38}) and a constant
\begin{equation}\label{A.42}
\int Dp_i Dq_i \int Dr_i Ds_i \exp \frac{-i}{2\alpha} \int dx \left( p_i r_i + p_is_i + q_ir_i\right).
\end{equation}
We then are left with
\begin{eqnarray}\label{A.43}
\tilde{Z} [j_i,k_i] &=& \int D\phi_i D\lambda_i \int D\bar{\xi}_i  D\bar{\zeta}_i \det F_{ij} \left(
\begin{array}{cc}
0 & H_{jk} \\
 H_{jk} & \lambda_\ell \frac{\partial H_{jk}}{\partial \phi_\ell}\end{array}\right)  \nonumber\\
& &\det G_{ij} \left(
\begin{array}{cc}
0 & H_{jk} \\
H_{jk} & \lambda_\ell \frac{\partial H_{jk}}{\partial \phi_\ell}\end{array}\right)\nonumber\\
& & \exp i \int dx \Bigg[ \mathcal{L}(\phi_i) + \lambda_k \frac{\partial \mathcal{L}(\phi_i)}{\partial \phi_k} + j_i\phi_i + k_i\lambda_i\nonumber \\
& - & \frac{1}{2\alpha}\Bigg( (F_{ij} \phi_j) \left( G_{ik} \left(\phi_k + H_{k\ell} \bar{\xi}_\ell\right)\right)\nonumber \\
&+& (F_{ij} \phi_j) \left( G_{ik}\left(\lambda_k + H_{k\ell} \bar{\zeta}_\ell + \lambda_m \frac{\partial H_{k\ell}}{\partial\phi_m}\bar{\xi}_\ell \right)\right)\nonumber\\
&+& \left(F_{ij} \lambda_j\right) \left( G_{ik} \left( \phi_k + H_{k\ell}\bar{\xi}_\ell\right)\right)\Bigg)\Bigg]
\end{eqnarray}
where $\bar{\xi}_i = \theta_i - \xi_i$, $\bar{\zeta}_i = \rho_i - \zeta_i$.

Exponentiation of the functional determinants in Eqs. (\ref{A.40}) and (\ref{A.43}) by use of
\begin{equation}\label{A.44}
\det M_{ij} = \int Dc_i D\bar{c}_i \exp \bar{c}_i M_{ij} c_j
\end{equation}
where $c_i$, $\bar{c}_i$ are Grassmann leads to Fermionic ghost fields. In Eq. (\ref{A.43}), $\bar{\xi}_i$ and $\bar{\zeta}_i$ are Bosonic ghost fields. We will not consider using the second condition of Eq.~\eqref{A.41} any further.

If we use
\begin{equation}\label{A.45}
\det \left( \begin{array}{cc}
0 & A \\
A & B\end{array}\right) = \det \left( \begin{array}{cc}
0 & A \\
A & A+B\end{array}\right)
\end{equation}
in Eq. (\ref{A.40}) and then use Eq. (\ref{A.44}), we find that
\begin{eqnarray}\label{A.46}
Z[j_i, k_i] &=& \int D\phi_i D\lambda_i \int DN_i DL_i \int Dc_i D\bar{c}_i Dd_i D\bar{d}_i\nonumber \\
& &\exp i \int dx \Bigg[ \mathcal{L}(\phi_i) + \lambda_k \frac{\partial\mathcal{L}(\phi_i)}{\partial\phi_k} + j_i \phi_i + k_i \lambda_i\nonumber\\
&+& \bar{c}_i F_{ij} H_{jk} d_k + \bar{d}_i F_{ij} H_{jk}  c_k
+ \bar{c}_i F_{ij} \left(H_{jk} + \lambda_\ell\frac{\partial H_{jk}}{\partial\phi_\ell} \right)c_k\nonumber \\
&+& \left( \frac{\alpha}{2} N_i N_i - N_iF_{ij} \left(\phi_j + \lambda_j\right) + \alpha N_i L_i - L_i F_{ij} \phi_j\right)\Bigg]
\end{eqnarray}
where $N_i$, $L_i$ are ``Nakanishi-Lautrup'' fields \cite{29}.

Provided $f_{mj\ell}$ in Eq. (\ref{A.36}) is independent of $\phi_i$, it can be shown that
\begin{eqnarray}\label{A.47}
\mathcal{L}(\phi_i) + \lambda_k \frac{\partial \mathcal{L}(\phi_i)}{\partial \phi_k} &+& \bar{c}_i F_{ij} H_{jk} d_k + \bar{d}_i F_{ij} H_{jk}c_k \nonumber\\
&+& \bar{c}_i F_{ij} \left( H_{jk} + \lambda_\ell \frac{\partial H_{jk}}{\partial\phi_\ell}\right)c_k\nonumber \\
&+& \left( \frac{\alpha}{2} N_i N_i - N_i F_{ij} (\phi_j + \lambda_j) + \alpha N_i L_i - L_i N_{ij} \phi_j\right)
\end{eqnarray}
is invariant under the transformation
\begin{equation}\label{A.48}
\delta \phi_i = H_{ij} c_j \epsilon
\end{equation}
\begin{equation}\label{A.49}
\delta \lambda_i = H_{ij} d_j \epsilon +\lambda_k \frac{\partial H_{ij}}{\partial\phi_k} c_k \epsilon
\end{equation}
\begin{equation}\label{A.50}
\delta N_i = 0 = \delta L_i
\end{equation}
\begin{equation}\label{A.51}
\delta \bar{c}_i = - \epsilon N_i
\end{equation}
\begin{equation}\label{A.52}
\delta \bar{d}_i = - \epsilon L_i
\end{equation}
\begin{equation}\label{A.53}
\delta c_i = - \frac{1}{2} f_{ijk} c_jc_k \epsilon
\end{equation}
\begin{equation}\label{A.54}
\delta d_i = - f_{ijk} c_j d_k \epsilon ,
\end{equation}
where $\epsilon$ is a Grassmann constant.  Eqs. (\ref{A.48}-\ref{A.54}) are the global ``BRST'' transformations associated with Eq. (\ref{A.47}) \cite{30}.  In Appendix \ref{appB} we will show that the presence of this invariance ensures that introduction of a Lagrange multiplier field as in Eq. (\ref{A.14}) for a gauge theory is consistent with unitarity, provided it is a nilpotent transformation \cite{31}.  This requires that $f_{ijk}$ is independent of $\phi_i$.

\section{The Scalar Field}

We will consider now a scalar field $\phi^a$ with the classical action
\begin{equation}\label{eq1}
S[\phi^a] =\int dx \left( \frac{1}{2} \left(\partial_\mu \phi^a\right)^2 - \frac{m^2}{2}(\phi^a)^2 - \frac{G}{4!}\left(\phi^a\phi^a\right)^2\right).
\end{equation}
It possesses a global $O(N)$ symmetry as well as the symmetry $\phi^a \rightarrow - \phi^a$.  The path integral quantization procedure leads to the generating functional
\begin{equation}\label{eq2}
Z[j^a] = \int D\phi^a \exp i \left( S(\phi^a) + \int dx j^a\phi^a\right).
\end{equation}
If we employ a background field $B^a$, then by Eqs. (\ref{A.7}, \ref{A.8}) the generating function for one particle irreducible diagrams is $\Gamma$ where
\begin{equation}\label{eq3}
e^{i\Gamma[B^a]} = \int DQ^a e^{i(S[B^a+Q^a)+j^aQ^a]}
\end{equation}
where now
\begin{equation}\label{eq4}
j^a = - \frac{\delta \Gamma[B^a]}{\delta B^a}.
\end{equation}
Divergences that arise in the course of performing a loop expression of Eq. (\ref{eq3}) can be absorbed into a renormalization of $B^a$, $m^2$ and $G$ when using dimensional regularization \cite{14}.

If now we introduce a LM field $\lambda^a$ to impose the equation of motion for $\phi^a$, our generating functional becomes, by Eq. (\ref{A.14})
\begin{eqnarray}\label{eq5}
Z[j^a,k^a] &=& \int  D\phi^a D\lambda^a \exp i \int dx \Big[ \frac{1}{2} \left(\partial_\mu\phi^a\right)^2 - \frac{m_1^2}{2}(\phi^a)^2 - 
\frac{G_1}{4!} (\phi^a\phi^a)^2\nonumber \\
& &- \lambda^a \left( \partial^2\phi^a + m_2^2 \phi^a + \frac{G_2}{3!} \phi^a\phi^b\phi^b\right)+ j^a\phi^a + k^a\lambda^a\Big].
\end{eqnarray}
In Eq. (\ref{eq5}) we distinguish $m_1^2$ and $m_2^2$ as well as $G_1$ and $G_2$ as the divergences arising in the one-loop contribution to $Z$ in Eq. (\ref{eq5}) renormalize $m_1^2$ and $G_1$, but not $m_2^2$ and $G_2$ \cite{4}.

To see this, we can first of all do a diagrammatic expansion of $Z$ in Eq. (\ref{eq5}).  Since the terms in the exponential of Eq. (\ref{eq5}) that are bilinear in $\phi^a$ and $\lambda^a$ are
\begin{equation}\label{eq6}
- \frac{1}{2} (\phi^a,\lambda^a)\left(
\begin{array}{cc}
\partial^2 + m_1^2 & \partial^2 + m_2^2\\
\partial^2 + m_2^2 & 0 \end{array}
\right)
\left(
\begin{array}{c}
\phi^a\\
\lambda^a\end{array}
\right)
\end{equation}
the propagators for $\phi^a$, $\lambda^a$ can be found from
\begin{equation}\label{eq7}
\left( \begin{array}{cc}
\partial^2 + m^2_1 & \partial^2 + m^2_2\\
     &   \\
\partial^2 + m^2_2 & 0 \end{array} \right)^{-1}
= \left(
\begin{array}{cc}
0 & \frac{1}{\partial^2 + m^2_2} \\
\frac{1}{\partial^2 + m^2_2} & 
-\frac{\partial^2 + m^2_1}{(\partial^2 + m^2_2)^2} \end{array} \right).
\end{equation}
So also, we have vertices $\phi$-$\phi$-$\phi$-$\phi$ and $\lambda$-$\phi$-$\phi$-$\phi$.
\begin{figure}[ht!]
\begin{align*}
\vcenter{\hbox{\includegraphics[scale=0.6]{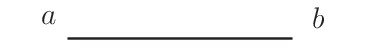}}} & = 0 \nonumber\\
& \\
\vcenter{\hbox{\includegraphics[scale=0.6]{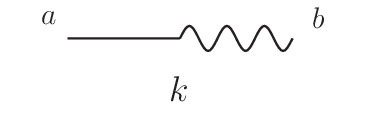}}}& = \frac{i}{k^2-m^2_2} \delta^{ab} \\
& \\
\vcenter{\hbox{\includegraphics[scale=0.6]{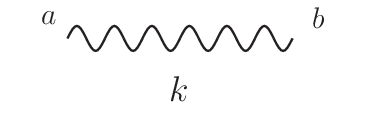}}} &=  -i \frac{k^2 - m^2_1}{(k^2-m^2_2)^2} \delta^{ab} \\
& \\
\vcenter{\hbox{\includegraphics[scale=0.6]{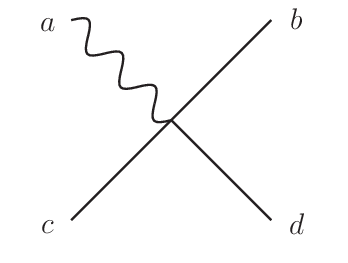}}}&= -iG_2 \left( \delta^{ab} \delta^{cd} + \delta^{ac} \delta^{bd} + \delta^{ad} \delta^{bc}\right) \\
& \\
\vcenter{\hbox{\includegraphics[scale=0.6]{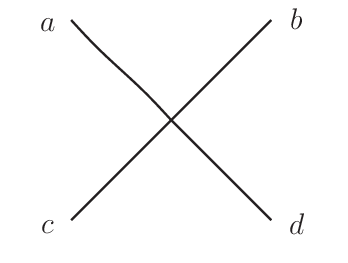}}}&= -iG_1 \left( \delta^{ab} \delta^{cd} + \delta^{ac} \delta^{bd} + \delta^{ad} \delta^{bc}\right) 
\end{align*}
\caption{Feynman rules of the Eq.~\eqref{eq5}. The solid and wavy lines represent the scalar field $ \phi^{a} $ and the LM field $ \lambda^{a} $, respectively.}
\end{figure}
With these Feynman rules, one cannot construct a Feynman diagram with more than one loop.  The divergent Feynman diagrams at one-loop order are in Fig. 2.
\begin{figure}[ht!]
    \centering
    \includegraphics[scale=0.7]{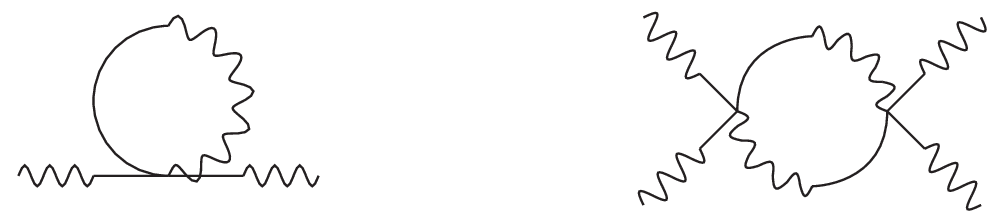}
    \caption{One-loop diagrams with non-amputated external legs.}
\end{figure}
The external legs have not been amputated.
The symmetry factor associated with these diagrams ensures that they are twice the corresponding diagram coming from Eq. (\ref{eq2}).

If we consider amputated one loop diagrams, as well as tree diagrams, then the two- and four-point functions are of the form (with $\epsilon = 2 - n/2$ in $n$-dimensions \cite{14}).
\begin{equation}\label{eq8}
\langle \phi^a\phi^b \rangle  = \int dx \left[ - \frac{1}{2} \phi^a \left( \partial^2 + m_1^2\right)\phi^a - \frac{G_2}{2}m_2^2 \phi^a\phi^a\left( \frac{\bar{c}}{\epsilon} + c_1 \ln \frac{m_2^2}{\mu^2}+ c_0\right)\right]
\end{equation}
\begin{equation}\label{eq9}
\langle \phi^a\phi^b\phi^c\phi^c \rangle  = \int dx \left[ - \frac{G_1}{4!} \left(\phi^a\phi^a\right)^2 - \frac{G_2^2}{4!} \left(\phi^a\phi^a\right)^2 \left( \frac{\bar{d}}{\epsilon} + d_1 \ln \frac{m_2^2}{\mu^2}+ d_0\right)\right]
\end{equation}
so that as $\epsilon \rightarrow 0$, all divergences are removed by renormalizing $m_1^2$ and $G_1$
\begin{equation}\label{eq10}
m_1^{2R} = m_1^2 + \frac{G_2\bar{c} m_2^2}{\epsilon}
\end{equation}
\begin{equation}\label{eq11}
G_1^R = G_1 + \frac{G_2^2\bar{d}}{\epsilon}.
\end{equation}
There is no need to renormalize the field $\phi^a$ \cite{14}.

These results can also be obtained by explicitly performing the functional integrals in Eq. (\ref{eq5}).  By use of Eq. (\ref{A.17}) we find that
\begin{eqnarray}\label{eq12}
Z\left[ j^a, k^a\right] &=& \sum_{\bar{\phi}^a}\exp i \int dx \left[ - \frac{1}{2} \bar{\phi}^a \left(\partial^2 + m_1^2\right) \bar{\phi}^a - 
\frac{G_1}{4!}\left( \bar{\phi}^a \bar{\phi}^a\right)^2 + j^a \bar{\phi}^a\right]\\
& & \textstyle{\det^{-1}} \left(\left(\partial^2 +m_2^2\right)\delta^{ab} + \frac{G_2}{3!} \left(\delta^{ab}\bar{\phi}^c \bar{\phi}^c
+ 2 \bar{\phi}^a\bar{\phi}^b\right)\right)\nonumber
\end{eqnarray}
where $\bar{\phi}^a$ satisfies 
\begin{equation}\label{eq13}
\left( \partial^2 + m_2^2\right) \bar{\phi}^a + \frac{G_2}{3!}\bar{\phi}^a\bar{\phi}^b\bar{\phi}^b = k^a .
\end{equation}

A perturbative expansion of $\bar{\phi}^a$ in powers of $G_2$ that follows from Eq. (\ref{eq13}) has a diagrammatic form given in fig. 3
\begin{figure}[ht]
    \centering
    \includegraphics[width=7in]{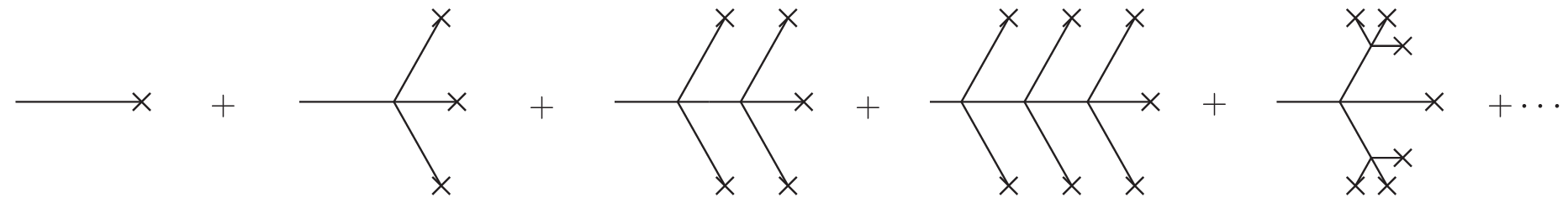}
    \caption{Diagrammatic form of the perturbative solution of Eq.~\eqref{eq13}.}
\end{figure}
where lines represent $(\partial^2 + m_2^2)^{-1}$, $ \--\!\!\!\--\!\!\!\!\times $ denotes a factor of $k^a$ and $ >\!<$ is a vertex associated with the coupling $G_2$.  The exponential in Eq. (\ref{eq12}) represents the sum of all tree-level Feynman diagrams and the functional determinant is the square of the contribution coming from one-loop Feynman diagrams such as those of fig. 2.

The elimination of divergences through Eqs. (\ref{eq10}) and (\ref{eq11}) results in having to use distinct masses and couplings for tree-level and one-loop diagrams, as given in the Feynman rules of Fig. 1.  This situation results in it not being possible to compute an $S$-matrix using the generating functional of Eq. (\ref{eq5}) in a way consistent with unitarity \cite{15}.  As a result, a Lagrange multiplier cannot be used to consistently eliminate diagrams beyond one-loop order when considering the scalar model of Eq. (\ref{eq1}).
Overcoming this problem by simply setting $m_1=m_2$ or $G_1=G_2$ is
inappropriate, as no symmetry relates $m_1$ to $m_2$ or $G_1$ to
$G_2$. (Gauge invariance imposes a relationship between analogous
parameters in gauge theories.)

We now will turn our attention to using a LM field in conjunction with YM theory.

\section{Yang-Mills Theory}

The second order YM action
\begin{equation}\label{eq14}
S_{2YM}[A] = \int dx \left(\frac{-1}{4} f_{\mu\nu}^a (A) f^{a\mu\nu} (A)\right)
\end{equation}
where
\begin{equation}\label{eq15}
f_{\mu\nu}^a (A) = \partial_\mu A_\nu^a - \partial_\nu A_\mu^a + gf^{abc} A^b_\mu A_\nu^c
\end{equation}
possesses the local gauge invariance
\begin{equation}\label{eq16}
\delta A_\mu^a = D_\mu^{ab} (A) \xi^b
\end{equation}
\begin{equation}\label{eq17}
\left( D_\mu^{ab} (A) \equiv \partial_\mu\delta^{ab} + gf^{apb} A_\mu^p\right).
\end{equation}

By introducing an auxiliary field $F_{\mu\nu}^a$ so that we have the first order YM action
\begin{equation}\label{eq18}
S_{1YM} [A, F] = \int dx \left( -\frac{1}{2} F_{\mu\nu}^a f^{a\mu\nu} (A) + \frac{1}{4} F_{\mu\nu}^a F^{a\mu\nu} \right)
\end{equation}
the interaction vertices for $S_{2YM}$ are simplified \cite{16,Frenkel:2017xvm,Lavrov:2021pqh}

We now consider the generating functional
\begin{eqnarray}\label{eq19}
Z_{1YM} \left[ j_\mu^a , J_{\mu\nu}^a \right] & =& \int DA_\mu^a DF_{\mu\nu}^a D\bar{c}^a Dc^a \exp i \int \Big( - \frac{1}{2} F_{\mu\nu}^a f^{a\mu\nu} + \frac{1}{4} F_{\mu\nu}^a F^{a\mu\nu}\nonumber \\
&+ & j_\mu^a A^{a\mu} + J_{\mu\nu}^a F^{a\mu\nu} - \frac{1}{2\alpha} (\partial \cdot A^a)^2 + \bar{c}^a \partial_\mu D^{ab\mu}c^b\Big)
\end{eqnarray}
where we have employed the gauge fixing
\begin{equation}\label{eq20}
\partial \cdot A^a = 0
\end{equation}
and have the ghost Lagrangian
\begin{equation}\label{eq21}
\mathcal{L}_{\mathrm{ghost}} = \bar{c}^a \partial \cdot D^{ab}(A) c^b
\end{equation}
which is the usual Faddeev-Popov ghost Lagrangian \cite{17}.

A background field $B_\mu^a$ will be introduced for $A_\mu^a$, and $G_{\mu\nu}^a$ for $F_{\mu\nu}^a$, following Eqs. (\ref{A.2}) to (\ref{A.8}).  If
\begin{equation}\label{eq22}
W_{1YM}\left[ j_\mu^a , J_{\mu\nu}^a\right] = -i\ln Z_{1YM} \left[ j_\mu^a , J_{\mu\nu}^a\right]
\end{equation}
and
\begin{equation}\label{eq23}
\Gamma_{1YM} \left[ A_\mu^a , G_{\mu\nu}^a\right] = W_{1YM} \left( j_\mu^a , J_{\mu\nu}^a\right)  - \int dx 
\left( j_\mu^a A^{a\mu} + J_{\mu}^a F^{a\mu\nu} \right)
\end{equation}
where
\begin{equation}\label{eq24}
B_\mu^a = \frac{\delta W_{1YM}}{\delta j^{a\mu}}
\end{equation}
\begin{equation}\label{eq25}
G_{\mu\nu}^a = \frac{\delta W_{1YM}}{\delta J^{a\mu\nu}}
\end{equation}
then the one-particle irreducible Feynman diagrams are generated by
\begin{eqnarray}\label{eq26}
\exp i \Gamma_{1YM}\left[B_\mu^a, G_{\mu\nu}^a \right] &=& \int
                                                           Dq_\mu^a
                                                           DQ_{\mu\nu}^a
                                                           D\bar{c} ^a
                                                           Dc^a 
 \exp i \int dx 
 \Big[ - \frac{1}{2} (G+Q)^a_{\mu\nu} f^{a\mu\nu} \left( B_\mu^a + q_{\mu}^a\right) 
\nonumber \\
&+& \frac{1}{4} \left( G_{\mu\nu}^a +  Q_{\mu\nu}^a\right) \left(G^{a\mu\nu} + Q^{a\mu\nu}\right)\nonumber \\
& + & j_\mu^a q^{a\mu} + J_{\mu\nu}^a Q^{a\mu\nu} - \frac{1}{2\alpha} \left(D^{ab}_\mu (B) q^{b\mu}\right)^2 + 
\bar{c}^a D_\mu^{ab} (B) D^{bc\mu} (B+q) c^a\Big].
\end{eqnarray}
Following refs. \cite{18}, we have used the gauge fixing
\begin{equation}\label{eq27}
D_\mu^{ab}(B) q^{b\mu} = 0
\end{equation}
in place of Eq. (\ref{eq20}).

We now will introduce LM fields to restrict radiative corrections to one-loop order.  If we start from the action $S_{2YM}$ of Eq. (\ref{eq14}), then this means considering
\begin{equation}\label{eq28}
S^\lambda_{2YM} [A, \lambda] = \int dx \left( - \frac{1}{4}\left( f_{\mu\nu}^a (A)\right)^2 + \lambda_\nu^a D_\mu^{ab} (A) f^{b\mu\nu} (A)\right).
\end{equation}
By Eqs. (\ref{A.32}) and (\ref{A.34}), the gauge invariance of Eq. (\ref{eq16}) is now accompanied by 
\begin{equation}\label{eq29}
\delta \lambda_\mu^a = gf^{abc} \lambda_\mu^b \xi^c
\end{equation}
as well as
\begin{equation}\label{eq30}
\delta \lambda_\mu^a = D_\mu^{ab} (A) \zeta^b.
\end{equation}
(In appendix \ref{appB} we will show how these gauge invariances lead to BRST
invariance of the effective Lagrangian that results from quantizing
this action using the Faddeev-Popov procedure, and demonstrate now
this invariance results in 
the cancellation of unphysical degrees of freedom so that
unitarity is maintained.)
We now consider the generating functional that is derived following Eq. (\ref{A.46})
\begin{eqnarray}\label{eq31}
Z_{2YM}^\lambda &=& \int DA_\mu^a D\lambda_\mu^a \int DN^a DL^a \int Dc^a D\bar{c}^a Dd^a D\bar{d}^a\nonumber\\
&&\exp i \int dx \Big[ - \frac{1}{4} f_{\mu\nu}^a (A) f^{a\mu\nu} (A)
   + \lambda_\nu^a D_\mu^{ab} (A) f^{b\mu\nu} (A) + j^a_\mu A^{a\mu} +
   k_\mu^a \lambda^{a\mu}  \\
&+& \bar{c}^a \partial \cdot D^{ab} (A) d^b + \bar{d}^a \partial \cdot D^{ab} (A) c^b + 
\bar{c}^a \partial \cdot D^{ab} (A+\lambda) c^b \nonumber \\
& + & \left( \frac{\alpha}{2}N^aN^a - N^a \partial \cdot (A^a + \lambda^a) + \alpha N^a L^a - L^a \partial \cdot A^a\right)\Big]\nonumber
\end{eqnarray}
if we accompany Eq. (\ref{eq20}) with the gauge condition
\begin{equation}\label{eq32}
\partial \cdot \lambda^a = 0.
\end{equation}
We now introduce background fields $B_\mu^a$ and $\Lambda_\mu^a$ for $A_\mu^a$ and $\lambda_\mu^a$ respectively so that $A_\mu^a = B_\mu^a + Q_\mu^a$ and $\lambda_\mu^a = \Lambda_\mu^a + q_\mu^a$.  If $\Gamma[B_{\mu ,}^a \Lambda_\mu^a]$ is the one-particle irreducible generating functional then 
\begin{eqnarray}\label{eq33}
e^{i\Gamma_{2YM}^\lambda[B,\Lambda]} &=& \int DQ_{\mu ,}^a Dq_\mu^a \det
 \left(\begin{array}{cc}
0 & D_\mu^{ab}(B)D^{bc\mu} (B+Q)\\
D^{ab}(B)D^{bc} (B+Q) & D^{ab}_\mu(B)D^{bc\mu} (B+Q+\Lambda + q)\end{array}
\right)\nonumber \\
& &\exp i \int dx \Big[ - \frac{1}{4} f_{\mu\nu}^a (B+Q) f^{a\mu\nu} (B+Q) + \left( \Lambda_\mu^a + q_\mu^a\right)
\left( D_\nu^{ab} (B+Q)f^{b\mu\nu}(B+Q)\right)\nonumber \\
&-& \frac{1}{2\alpha} \left( D_\mu^{ab} (B)Q^{b\mu}\right)^2 -
    \frac{1}{\alpha}\left( D_\mu^{ab} (B)Q^{b\mu}\right)\left(
    D_\nu^{ac} (B)q^{c\nu}\right)
 + j_\mu^a Q^{a\mu} + k_\mu^a q^{a\mu}\Big]
\end{eqnarray}
with the gauge conditions
\begin{equation}\label{eq34}
D^{ab}_\mu (B) Q^{b\mu} = 0 = D_\mu^{ab} (B) q^{b\mu}.
\end{equation}
With this gauge fixing we have maintained the ``background gauge invariance''
\begin{subequations}\label{eq35}
\begin{eqnarray}
\delta B^a_\mu &=& D^{ab}_\mu (B) \xi^b \\
\delta Q^{a}_{\mu} &=& gf^{abc}Q^b_\mu\xi^c\\
\delta \Lambda^{a}_\mu &=& gf^{abc}\Lambda^a_\mu \xi^c\\
\delta q^a_\mu &=& gf^{abc}q^a_\mu \xi^c\\
\delta \Lambda^{a}_\mu &=& D^{ab}_\mu (B) \zeta^b\\
\delta q^{a}_\mu &=& gf^{abc} Q^b_\mu  \zeta^c \\
\end{eqnarray}
\end{subequations}
but have broken the gauge invariances
\begin{subequations}\label{eq36}
\begin{eqnarray}
\delta B_\mu^a &=& 0\\
\delta Q_\mu^a &=& D_\mu^{ab} (B+Q) \xi^b\\
\delta \Lambda_\mu^a &=& 0\\
\delta q_\mu^a &=& gf^{abc} \left(\Lambda_\mu^b + q_\mu^b\right) \xi^c\\
\delta q_\mu^a &=& D_\mu^{ab} (B+Q) \zeta^b\\
\end{eqnarray}
\end{subequations}
that were present in $S_{2YM}^\lambda (B+Q, \Lambda + q)$ of Eq. (\ref{eq28}).

One could find the Feynman rules associated with Eq. (\ref{eq33}) and perform a diagrammatic expansion of $\Gamma_{2YM}^\lambda$.  However, it is possible perform all the functional integrals that occur, and following the steps that lead to Eq. (\ref{A.26}) we obtain (since 
$\det\left( \begin{array}{cc} 0 &A\\A&B\end{array}\right) = \det \left(\begin{array}{cc} 0 &A\\A&A+B\end{array}\right) = \det^2A$)
\begin{eqnarray}\label{eq37}
 \displaystyle{e^{i\Gamma_{2YM}^\lambda [B_\mu^a,\Lambda_\mu^a]}}
&=& \sum_{\bar{Q}_\mu^a} \exp i \int dx \Big[ - \frac{1}{4} f_{\mu\nu}^a (B+\bar{Q}) f^{a\mu\nu} (B+\bar{Q})\\
& -& \frac{1}{2\alpha} \left( D^{ab} (B) \cdot \bar{Q}^b\right)^2 + \Lambda_\mu^a \left( D_\nu^{ab} (B+\bar{Q}) f^{b\mu\nu} (B+\bar{Q})\right) 
+ j_\mu^a \bar{Q}^{a\mu}\Big] \det{}^2 \left(  D_\mu^{ab} (B) D^{bc\mu} (B+\bar{Q})\right)\nonumber\\
& & \times \det{}^{-1} \Big\{ \frac{1}{2} \left( D^{2ab} (B)\eta_{\nu\mu} + D^{2ba}(B)\eta_{\nu\mu}\right) - \frac{1}{2}\left(1 - \frac{1}{\alpha}\right)\left( D_\mu^{ap} (B) D_\nu^{pb} (B) + D_\nu^{bp} (B) D_\mu^{pa} (B)\right)\nonumber \\
&+& \left( f^{apb} f_{\mu\nu}^p (B)+ f^{bpa} f_{\nu\mu}^p (B)\right) + \frac{1}{2} \Big[ D_\lambda^{ap} (B) \left( gf^{pqb} \bar{Q}^q_\lambda\right)\eta_{\mu\nu}\nonumber + D_\lambda^{bp} (B) \left( gf^{pqa} \bar{Q}^q_\lambda\right)\eta_{\nu\mu}\Big] \nonumber \\
&+& \frac{1}{2} \left[ D^{ap}_\nu (B) \left( gf^{pbq} \bar{Q}^q_\mu\right)  + D_\mu^{bp}(B) \left( gf^{paq} \bar{Q}^q_\nu\right)\right]
- \frac{1}{2} \left[\left( D^{pq}_\mu (B)\bar{Q}^q_\nu\right) \left( gf^{pab}\right) + \left( D_\nu^{pq}(B)\bar{Q}^q_\mu\right) 
\left( gf^{pba}\right)\right]\nonumber \\
&+& \frac{1}{2} \left[ D^{bp}_\mu (B) \left( gf^{par} \bar{Q}^r_\nu\right)  + D_\nu^{ap}(B) \left( gf^{pbr} \bar{Q}^r_\mu\right)\right]
-\frac{1}{2} \left[\left( D^{pq}_\nu (B)\bar{Q}^q_\mu\right) \left( gf^{pba}\right) + \left( D_\mu^{pq}(B)\bar{Q}^q_\nu\right) 
\left( gf^{pab}\right)\right]\nonumber \\
&+& \frac{1}{2} \left[ \eta_{\mu\nu}D^{bq}_\lambda (B) \left( gf^{pqa} \bar{Q}^q_\lambda\right) + \eta_{\nu\mu} D_\lambda^{aq}(B) \left( gf^{pqb} \bar{Q}^q_\lambda \right)\right]
-\frac{g^2}{2} \Big[ f^{mab} f^{mrs} \bar{Q}^r_\mu \bar{Q}^s_\nu +   f^{mba} f^{mrs} \bar{Q}^r_\nu \bar{Q}^s_\mu \nonumber \\
&+&  f^{maq} f^{mbs} \bar{Q}^q_\lambda \eta_{\mu\nu} \bar{Q}^s_\lambda + f^{mbq} f^{mas} \bar{Q}^q_\lambda 
\eta_{\nu\mu} \bar{Q}^s_\lambda 
+  f^{maq} f^{msb} \bar{Q}^q_\nu \bar{Q}^s_\mu + f^{mbq} f^{msa} \bar{Q}^q_\mu
 \bar{Q}^s_\nu \Big] \Big\}\nonumber
\end{eqnarray}
where $\bar{Q}_\mu^a$ satisfies
\begin{eqnarray}\label{eq38}
D^{2ab} (B) \bar{Q}_\mu^b &-& \left(1-\frac{1}{\alpha}\right) D_\mu^{ap} (B) D^{pb}_\lambda (B) \bar{Q}^{b\lambda} + 2gf^{abc}f_{\mu\nu}^b (B) \bar{Q}^{c\nu}\nonumber \\
&+& D_\lambda^{ab} (B) \left(gf^{bpq} \bar{Q}^{p\lambda}\bar{Q}^{q}_\mu \right) - g 
\left( D_\mu^{pq} (B) \bar{Q}_\lambda^q\right)\left(f^{paq}\bar{Q}^{q\lambda}\right)
\nonumber \\
& -&g\left( D_\lambda^{pq} (B) \bar{Q}^q_\mu\right) \left(f^{pqa} \bar{Q}^{q\lambda}\right) - g^2 f^{maq} f^{mrs} 
\bar{Q}_\lambda^q \bar{Q}_\mu^r \bar{Q}^{s\lambda} = k^a_\mu .
\end{eqnarray}
If $k_\mu^a = \bar{Q}_\mu^a = 0$, then Eq. (\ref{eq37}) reduces to
\begin{eqnarray}\label{eq39}
e^{i\Gamma_{2YM}^\lambda [B,\Lambda]} &=& \exp i \int dx \left[ -\frac{1}{4} f_{\mu\nu}^a (B) f^{a\mu\nu} (B) + \Lambda_\mu^a
D^{ab}_\nu (B) f^{b\mu\nu}(B)\right]\nonumber \\
&&\hspace{-1cm} \det{}^2 \left( D_\mu^{ab}(B) D^{bc\mu}(B) \right) \det{}^{-1} \left( D^{2ab}(B) \eta_{\mu\nu} - \left( 1 - \frac{1}{\alpha}\right) D_\mu^{ap} (B) D_\nu^{pb} (B) + 2g f^{apb} f_{\mu\nu}^p (B)\right).
\end{eqnarray}
The exponential in Eq. (\ref{eq39}) is the sum of all tree diagrams and the determinants are the sum of all one-loop diagrams contributing to 
$\Gamma_{2YM}^\lambda$; no higher loop contributions occur.  As noted
in the appendix, the equation of motion for the background field 
$B_\mu^a$
(Eq. (\ref{A.12})) does not have to be satisfied. Since the LM does not appear as an external state, we set $ \Lambda_{\mu}^{a} =0$.

We now consider renormalization of $g$ and $B_\mu^a$ needed to remove divergences arising in the perturbative evaluation of the functional determinants appearing in Eq. (\ref{eq39}) \cite{4}.  Maintaining the gauge invariance of Eq. (\ref{eq35}) ensures that the product $g B_\mu^a$ is invariant under renormalization, so that the renormalized coupling $g_R$ and renormalized field $B_{R\mu}^a$ satisfy \cite{18}
\begin{equation}\label{eq40}
g_R B_{R\mu}^a = g B_\mu^a .
\end{equation}
Consequently, if the two-and three-point functions determine the renormalization of $g$ and $B_\mu^a$, we use the result that
\begin{equation}\label{eq41}
\langle BB\rangle  = \frac{1}{2} B_\mu^a \left( p^2 \eta^{\mu\nu} - p^\mu p^\nu/p^2\right) B_\nu^a \left[1 + g^2 \left( \frac{\bar{D}}{\epsilon} + D_1 \ln \frac{p^2}{\mu^2} + D_0\right)\right]
\end{equation}
and
\begin{equation}\label{eq42}
\langle BBB\rangle  = f^{abc} B_\mu^a B_\nu^b B_\lambda^c V^{\mu\nu\lambda}\left[g + g^3 \left( \frac{\bar{E}}{\epsilon} + E_1 \ln \frac{p^2}{\mu^2} + E_0\right)\right].
\end{equation}
One does not have separate couplings and fields at tree and one loop
order, as with scalar fields in Eq. (\ref{eq5}), in order to eliminate
divergences occurring in Eqs. (\ref{eq41}, \ref{eq42}) as $\epsilon \rightarrow 0$.  One defines in this case
\begin{equation}\label{eq43}
B_{R_{\mu}}^a B_{R\nu}^b = B_\mu^a B_\nu^b \left( 1 + \frac{\bar{g}^2\bar{D}}{\epsilon}\right)
\end{equation}
\begin{equation}\label{eq44}
B_{R_{\mu}}^a B_{R_{\nu}}^b B_{R_\lambda}^c g_R = B_\mu^a B_\nu^b  B_\lambda^c \left( g + \frac{\bar{g}^3\bar{E}}{\epsilon}\right) .
\end{equation}
By Eq. (\ref{eq43})
\begin{equation}\label{eq45}
B_{R_{\mu}}^a = B_\mu^a \left( 1 + \frac{\bar{g}^2\bar{D}}{\epsilon}\right)^{1/2}
\end{equation}
and so by Eq. (\ref{eq44})
\begin{equation}\label{eq46}
 g_R = \left( g + \frac{\bar{g}^3\bar{E}}{\epsilon}\right)\left( 1 + \frac{\bar{g}^2\bar{D}}{\epsilon}\right)^{-3/2} .
\end{equation}
In order that Eq. (\ref{eq40}) is satisfied, we must have
\begin{equation}\label{eq47}
\bar{D} = \bar{E},
\end{equation}
and so by Eq. (\ref{eq46})
\begin{equation}\label{eq48}
 g_R =  g \left( 1 + \frac{\bar{g}^2\bar{E}}{\epsilon}\right)^{-1/2}.
\end{equation}
Eqs. (\ref{eq41}) and (\ref{eq42}) now become
\begin{equation}\label{eq49}
\langle BB\rangle  = \frac{1}{2} B_{R_{\mu}}^a \left(p^2 \eta^{\mu\nu} - p^\mu p^\nu / p^2\right) 
B_{R_{\nu}}^a \left[ 1 + g_R^2 \left( D_1 \ln \frac{p^2}{\mu^2} + D_0\right)\right]
\end{equation}
\begin{equation}\label{eq50}
\langle BBB\rangle  = f^{abc} B_{R_{\mu}}^a B_{R_{\nu}}^b B_{R_{\lambda}}^c V^{\mu\nu\lambda} 
\left[ g_R  + g_R^3 \left( E_1 \ln \frac{p^2}{\mu^2} + E_0\right)\right].
\end{equation}
Dimensionally,
\begin{equation}\label{eq51}
g = g_0\mu^\epsilon
\end{equation}
where $\mu$ is an arbitrary dimensionful constant so Eq. (\ref{eq48}) implies that
\begin{equation}\label{eq52}
\mu \frac{\partial g_R}{\partial_\mu} = \epsilon g_R - \bar{E} g_R^3 .
\end{equation}
As $\epsilon \rightarrow 0$, we arrive at the exact results for the $\beta$ function associated with $g$,
\begin{equation}\label{eq53}
\beta (g_R) = -\bar{E} g_R^3
\end{equation}
which is twice the usual one-loop result for the $\beta$-function in YM theory \cite{19}.  Since $gB_\mu^a$ is independent of $\mu$, Eqs. (\ref{eq40}) and (\ref{eq53}) show that
\begin{equation}\label{eq54}
\mu \frac{\partial B_{R\mu}^a}{\partial \mu} = \bar{E} g_R^2 B_{R\mu}^a .
\end{equation}

Having established how YM theory with a LM field is renormalized, we now consider how a matter field, in the form of a scalar field $\phi^a$, is coupled to $A^a_\mu$.  This involves supplementing $S_{2YM}$ in Eq. (\ref{eq14}) with
\begin{equation}\label{eq55}
S_{A\phi} = \int dx \left[ \frac{1}{2} \left( D_\mu^{ab} (A) \phi^b \right)^2 - \frac{m^2}{2} \phi^a \phi^a - \frac{G}{4!} \left( \phi^a\phi^a\right)^2\right]
\end{equation}
where $D_\mu^{ab}(A)$ is defined in Eq. (\ref{eq17}).  If now we were to make use of LM fields to eliminate diagrams beyond one-loop order, we would be using the classical action
\begin{eqnarray}\label{eq56}
S_{A\phi}^{\lambda\sigma} &=& \int dx \Big[ - \frac{1}{4} f_{\mu\nu}^a (A) f^{a\mu\nu} (A) + \frac{1}{2} \left( D_\mu^{ab} (A) \phi^b\right)^2 - \frac{m_1^2}{2!} \phi^a \phi^a \nonumber \\
& -& \frac{G_1}{4!} (\phi^a\phi^a)^2 + \lambda_\nu^a \left( D_\mu^{ab} (A) f^{b\mu\nu} (A) + gf^{apq} \phi^p D^{qr\nu} (A)\phi^r\right)\nonumber \\
& -&\sigma^a \left( D_\mu^{ab} (A) D^{bc\mu} (A) \phi^c + m_2^2 \phi^a + \frac{G_2}{3!}\phi^a\phi^b\phi^b\right)\Big] .
\end{eqnarray}
This action is invariant under the gauge transformations of Eqs. (\ref{eq16}), (\ref{eq29}) and (\ref{eq30}) along with (using Eqs. (\ref{A.32}) and (\ref{A.34})
\begin{equation}\label{eq57}
\delta\phi^a = gf^{abc} \phi^b\xi^c
\end{equation}
\begin{equation}\label{eq58}
\delta \sigma^a = gf^{abc} \sigma^b \xi^c
\end{equation}
and
\begin{equation}\label{eq59}
\delta\sigma^a = gf^{abc} \phi^b \zeta^c.
\end{equation}
As in Eq. (\ref{eq5}), renormalization of divergences arising at one-loop order in radiative effects involving scalars makes it necessary to have distinct masses and couplings at tree-level order ($m_1^2$ and $G_1$) and one-loop order ($m_2^2$ and $G_2$).  This again leads to results inconsistent with unitarity, and so all terms in Eq. (\ref{eq56}) proportional to $\sigma^a$ are to be discarded.

However, eliminating the term
\begin{equation}\label{eq60}
\sigma^a D_\mu^{ab} (A) D^{bc\mu} (A) \phi^c
\end{equation}
from Eq. (\ref{eq56}) breaks the invariance of Eq. (\ref{eq30}) unless the term
\begin{equation}\label{eq61}
\lambda_\nu^a \left( gf^{apq} \phi^p D^{qr\nu} (A) \phi^r\right)
\end{equation}
is also removed.  We are then left with the action
\begin{eqnarray}\label{eq62}
S_{A\phi}^\lambda &=& \int dx \Big[ - \frac{1}{4} f_{\mu\nu}^a (A) f^{a\mu\nu} (A) + \frac{1}{2} \left( D_\mu^{ab} (A) \phi^b\right)^2 - 
\frac{m^2}{2}\phi^a\phi^a \nonumber \\
&-& \frac{G}{4!} (\phi^a\phi^a)^2 + \lambda_\nu^a D_\mu^{ab}(A) f^{b\mu\nu}(A)\Big].
\end{eqnarray}
This is invariant under the gauge transformations of Eqs. (\ref{eq16}, \ref{eq29}, \ref{eq30}, \ref{eq57}).

The generating functional is, by Eq. (\ref{A.46}),
\begin{eqnarray}\label{eq63}
Z_{A\phi}^\lambda \left[ j_\mu^a , k_\mu^a , J^a\right] &=& \int D\phi^a DA_\mu^a D\lambda_\mu^a \int Dc^a D\bar{c}^a Dd^a D\bar{d}^a\nonumber\\
& &\exp i \Big[ S_{A\phi}^\lambda + \int dx \Big( \bar{c}^a \partial \cdot D^{ab} (A + \lambda)c^b\nonumber \\
&+& \bar{d}^a  \partial \cdot D^{ab} (A) c^b +  \bar{c}  \nonumber^a \partial \cdot D^{ab}(A) d^b 
- \frac{1}{2\alpha} (\partial \cdot A^a)^2 - \frac{1}{\alpha} \partial \cdot A^a \partial \cdot \lambda^a\nonumber \\
&+& j_\mu^a A^{a\mu} + k_\mu^a \lambda^{a\mu} + J^a \phi^a\Big)\Big]
\end{eqnarray}
if we use the gauge fixing of Eqs. (\ref{eq20}) and (\ref{eq32}).

The fields $\phi^a$, $A_\mu^a$ and $\lambda_\mu^a$ are expanded about backgrounds $\Phi^a$, $B_\mu^a$, $\Lambda_\mu^a$ so that
\begin{equation}\label{eq64}
\phi^a = \Phi^a + \psi^a
\end{equation}
\begin{equation}\label{eq65}
A^a_\mu = B^a_\mu + Q^a_\mu
\end{equation}
\begin{equation}\label{eq66}
\lambda^a_\mu = \Lambda^a_\mu + q^a_\mu .
\end{equation}
Using the same steps used to arrive at Eq. (\ref{eq39}) when there is no scalar ``matter'' field $\phi^a$, we find that
\begin{eqnarray}\label{eq67}
e^{i\Gamma_{A\phi}^\lambda [B,\Lambda , \Phi]} &=&  \int D\psi^a \exp i \int dx 
\Big[ -\frac{1}{4} f_{\mu\nu}^a (B) f^{a\mu\nu} (B) + \Lambda_\mu^a
D^{ab}_\nu (B) f^{b\mu\nu}(B) \nonumber \\
&+& \frac{1}{2} \left( D_\mu^{ab}(B) \left( \Phi^b + \psi^b\right)\right)^2 - \frac{m^2}{2} \left( \Phi^a + \psi^a\right)
\left( \Phi^a + \psi^a\right)\nonumber \\
&-& \frac{G}{4!} \left( \left( \Phi^a + \psi^a\right)\left( \Phi^a + \psi^a\right)\right)^2\Big] \det{}^2 \left(D_\mu^{ab}(B) D^{bc\mu} (B)\right)\nonumber \\
& &\det{}^{-1} \Big( D^{2ab} (B) \eta_{\mu\nu} - (1-\frac{1}{\alpha}) D_\mu^{ap} (B) D^{pb}_\nu(B) + 2gf^{apb} f_{\mu\nu}^p (B)\Big).
\end{eqnarray}
From Eq. (\ref{eq67}) it follows that a perturbative expansion of $\Gamma_{A\phi}^\lambda[B, \Lambda , \Phi]$ has the following contributions:
\begin{enumerate}
\item loops involving scalar fields $\phi^a$ propagating in the presence of a background scalar field $\Phi^a$ and a background vector gauge field $B_\mu^a$
\item all tree level diagrams involving the vector gauge field, given by the exponential\\
 $\exp i \int dx \left( - \frac{1}{4} f_{\mu\nu}^a (B) f^{a\mu\nu}(B)\right)$ \cite{20}.
\item twice the contribution of all one-loop diagrams arising in normal YM theory when a background field is used, coming from the functional determinants in Eq. (\ref{eq67}).
\item no higher loop contributions involving only the propagation of the gauge field.
\end{enumerate}

The terms in Eq. \eqref{eq67} that exclusively involve the background fields are
\be\label{e455}
\exp i \int dx\left[-\frac 1 4 f^a_{\mu\nu}(B)f^{a \mu\nu}(B)
+ \frac 1 2 (D^{ab}_{\mu}(B)\Phi^b)^2 - \frac{m^2}{2} \Phi^a \Phi^a
-\frac{G}{4!}(\Phi^a\Phi^a)^2 
+ \Lambda^a_\mu D^{ab}_\nu(B) f^{b\mu\nu}(B)\right].
\ee
If we set the field $\Lambda^a_\mu$ equal to zero, 
the remaining terms in the exponential in \eqref{e455} give the tree-level diagrams that follow from the classical action for a YM field coupled to a scalar. Alternatively, one could keep $\Lambda^a_\mu$ in the exponential and supplement Eq. \eqref{e455} with the extra term
\be\label{e456}
\exp i \int dx \left[\Lambda^a_\mu\left(f^{abc} \Phi^b D^{cd\mu}(B) \Phi^d\right)\right]. 
\ee
In this case, Eqs. \eqref{e455} and \eqref{e456} together are gauge invariant and are consistent with having tree diagrams follow from the classical action for a YM field coupled to a scalar with a LM field ensuring that the classical equations of motion for the YM field are satisfied.

As a result of this, the renormalization group functions for $g_R$ and $B_{R\mu}^a$ in Eqs. (\ref{eq53},\ref{eq54}) can receive loop contributions of higher order in $G_R$.  So also, the usual renormalization group functions for $G_R$, $m_R^2$ and $\phi_R$ are unaltered.

We now will consider how the arguments leading to Eq. (\ref{eq67}) can be used when there is a gravitational field interacting with a scalar field.

\section{Gravity}

At one-loop order, the EH action is renormalizable as divergences
vanish when the metric $g_{\mu\nu}$ satisfies the classical equations
of motion \cite{1}.  However, at higher loop order \cite{7}, or if the
metric couples to matter fields \cite{1,8,9}, then renormalizability
is lost
as divergences no longer disappear when equations of motion are satisfied.  
This has made it interesting to examine the consequences of using a LM field to limit radiative corrections to the EH action to one-loop order \cite{21,21a}.  In this section we will further consider use of a LM field in conjunction with the EH action supplemented by matter fields which couple to the metric. Our matter field will be a self-interacting scalar field.

The first-order form of the EH action is useful, as in this form, the EH action has only a three-point vertex \cite{22}.  Here however we will use the second order form of the EH action, treating the metric $g_{\mu\nu}$ as a gauge field.  This action is
\begin{equation}\label{eq68}
S_{EH} = \int dx \left( - \frac{1}{\kappa^2} \sqrt{-g} \;\;g^{\mu\nu} R_{\mu\nu} (\Gamma)\right)
\end{equation}
where $\kappa^2 = 16\pi G_N$, $g = \det g_{\mu\nu}$, $\Gamma_{\mu\nu}^\lambda = \frac{1}{2} g^{\lambda\rho}\left(g_{\mu\rho ,\nu} +
g_{\nu\rho ,\mu} - g_{\mu\nu ,\rho}\right)$ and\\
$R_{\mu\nu} = - \left( \Gamma_{\mu\nu , \lambda}^\lambda - \Gamma_{\mu\lambda , \nu}^\lambda + \Gamma_{\mu\nu}^\lambda \Gamma_{\lambda\sigma}^\sigma - \Gamma_{\mu\lambda}^\sigma \Gamma_{\nu\sigma}^\lambda \right)$. The metric is coupled to a scalar matter field $\phi$ with the action
\begin{equation}\label{eq69}
S_{g\phi} = \int dx \left( \sqrt{-g}\right)\left( - \frac{1}{2} g^{\mu\nu} \partial_\mu \phi \partial_\nu \phi - \frac{m^2}{2} \phi^2 - \frac{G}{4!} \phi^4 \right).
\end{equation}
Due to diffeomorphism invariance, $S_{EH} + S_{g\phi}$ is invariant under the infinitesimal transformation
\begin{equation}\label{eq70}
\delta g_{\mu\nu} = g_{\mu\alpha} \xi_{,\nu}^\alpha + g_{\nu\alpha} \xi_{, \mu}^\alpha + \xi^\alpha g_{\mu\nu ,\alpha}
\end{equation}
\begin{equation}\label{eq71}
\delta \phi = \xi^\alpha \phi_{, \alpha}.
\end{equation}

The nature of the EH action makes it impossible to perform any perturbative expansion of the generating functional without using a background field $\bar{g}_{\mu\nu}$ for the metric $g_{\mu\nu ,}$ so that \cite{23}
\begin{equation}\label{eq72}
g_{\mu\nu} = \bar{g}_{\mu\nu} + \kappa h_{\mu\nu} .
\end{equation}
Quite often, this background field is chosen to be flat
\begin{equation}\label{eq73}
\bar{g}_{\mu\nu} = \eta_{\mu\nu}.
\end{equation}
An expansion of the geometric quantities found in Eqs. (\ref{eq68}) and (\ref{eq69}) in powers of $h_{\mu\nu}$ is found in ref. \cite{1}.

If $V_{\alpha ; \bar{\beta}}$ denotes a covariant derivate of $V_\alpha$ using the background field $\bar{g}_{\mu\nu}$, then the gauge transformation of Eq. (\ref{eq70}) can be written as
\begin{equation}\label{eq74}
\delta \bar{g}_{\mu\nu} = 0
\end{equation}
\begin{equation}\label{eq75}
\delta h_{\mu\nu} = \frac{1}{\kappa} \left( \xi_{\mu ;\bar{\nu}} + \xi_{\nu ;\bar{\mu}} \right) 
+ \xi^\lambda 
h_{\mu\nu ;\bar{\lambda}} + h_{\mu\lambda} \xi^\lambda_{\;\;; \bar{\nu}} +  h_{\nu\lambda} \xi^\lambda_{\;\;; \bar{\mu}}.
\end{equation}
Indices are raised and lowered using $\bar{g}_{\mu\nu}$.

Upon varying $g_{\mu\nu}$ in $S_{EH} + S_{g\phi}$, we find that
\begin{equation}\label{eq76}
\delta \left( S_{EH} + S_{g\phi} \right) = \int dx \sqrt{-g} \delta g_{\mu\nu} \left[ \frac{+1}{\kappa^2} \left( R^{\mu\nu} - \frac{1}{2} g^{\mu\nu} R\right) + \frac{1}{2} T^{\mu\nu}\right]
\end{equation}
where
\begin{equation}\label{eq77}
T^{\mu\nu} = g^{\mu\alpha} g^{\nu\beta} \phi_{,\alpha}\phi_{,\beta} - g^{\mu\nu} \left( \frac{1}{2} g^{\alpha\beta} \phi_{,\alpha} \phi_{,\beta} + \frac{m^2}{2} \phi^2 + \frac{G}{4!} \phi^4\right).
\end{equation}

If a LM field $\lambda^{\mu\nu}$ is used to impose the equation of motion for $g_{\mu\nu}$ that follows from $S_{EH}$, we have the action
\begin{equation}\label{eq78}
S_{EH}^\lambda = \frac{1}{\kappa^2} \int dx \sqrt{-g} \left[ -g^{\mu\nu} R_{\mu\nu} + \lambda^{\mu\nu} \left( R_{\mu\nu} - \frac{1}{2} g_{\mu\nu} R\right)\right].
\end{equation}

It now is possible to follow the steps which led to the generating functional $\Gamma_{2YM}^\lambda$ in Eq. (\ref{eq33}).  If we have background fields $\bar{g}_{\mu\nu}$ and $\bar{\lambda}^{\mu\nu}$ for $g_{\mu\nu}$ and $\lambda^{\mu\nu}$ so that 
\begin{equation}\label{eq79}
g_{\mu\nu} = \bar{g}_{\mu\nu} + \kappa h_{\mu\nu}
\end{equation}
\begin{equation}\label{eq80}
\lambda_{\mu\nu} = \bar{\lambda}_{\mu\nu} + \kappa \sigma_{\mu\nu}
\end{equation}
then $S_{EH}^\lambda$ in Eq. (\ref{eq77}) is invariant under the gauge transformation of Eq. (\ref{eq70}) combined with
\begin{equation}\label{eq81}
\delta \lambda_{\mu\nu} = \lambda_{\mu\alpha} \xi_{,\nu}^\alpha + \lambda_{\nu\alpha} \xi_{,\mu}^\alpha + \xi^\alpha 
\lambda_{\mu\nu , \alpha}
\end{equation} 
and
\begin{equation}\label{eq82}
\delta \lambda_{\mu\nu} = \left( g_{\mu\alpha} \zeta_{,\nu}^\alpha + g_{\nu\alpha} \zeta_{,\mu}^\alpha + \zeta^\alpha 
g_{\mu\nu , \alpha}\right)
\end{equation}
as can be seen from Eqs. (\ref{A.32}) and (\ref{A.34}).  Eq. (\ref{eq81}) follows from the fact that $\lambda_{\mu\nu}$ is a tensor under a diffeomorphism transformation while Eq. (\ref{eq82}) follows from the fact the $S_{EH}$ in Eq. (\ref{eq68}) is invariant under a diffeomorphism transformation.

Just as Eqs. (\ref{eq74}) and (\ref{eq75}) follow from Eq. (\ref{eq70}), we see that Eqs. (\ref{eq81}) and (\ref{eq82}) result in
\begin{equation}\label{eq83}
\delta \bar{\lambda}_{\mu\nu} = 0
\end{equation}
\begin{equation}\label{eq84}
\delta \sigma_{\mu\nu} = \left( \frac{1}{\kappa} \bar{\lambda}_{\mu\alpha} + \sigma_{\mu\alpha}\right) \xi_{; \bar{\nu}}^\alpha
+ \left( \frac{1}{\kappa} \bar{\lambda}_{\nu\alpha} + \sigma_{\nu\alpha}\right) \xi_{; \bar{\nu}}^\alpha + 
\xi^\alpha \left(\frac{1}{\kappa} \bar{\lambda}_{\mu\nu} +
  \sigma_{\mu\nu}\right)_{; {\bar \alpha}}.
\end{equation}
\begin{equation}\label{eq85}
\delta \sigma_{\mu\nu} =  \frac{1}{\kappa}\left( \zeta_{\mu ; \bar{\nu}} + \zeta_{\nu ; \bar{\mu}}\right)
+ \zeta^\lambda h_{\mu\nu ; \bar{\lambda}} + 
h_{\mu\lambda} \zeta^\lambda_{; \bar{\nu}} + h_{\nu\lambda} \zeta^\lambda_{;\bar{\mu}}.
\end{equation}

It is now possible to break the invariance of Eqs. (\ref{eq74},
\ref{eq75}, \ref{eq83}, \ref{eq84}, \ref{eq85}) by the gauge fixing conditions
\begin{equation}\label{eq86}
h_{\mu\nu ;}^{\;\;\;\;\bar{\nu}} - k h_{\nu\;\; ;\bar{\mu}}^{\;\;\nu} = 0 = \sigma_{\mu\nu ;}^{\;\;\;\;\bar{\nu}} - 
k\sigma_{\nu\;\; ; \bar{\mu}}^\nu .
\end{equation}
We now can follow the steps used to find the generating functional $\Gamma_{g\phi}^\lambda \left[\bar{\lambda}, \bar{g}, \bar{\Phi}\right]$ for one-particle irreducible graphs when a scalar $\phi = \Phi + \psi$ with background $\Phi$ is in the presence of  background $\bar{g}_{\mu\nu}$ and $\bar{\lambda}_{\mu\nu}$.  In analogy with Eq. (\ref{eq67}), this leads to
\begin{eqnarray}\label{eq87}
\exp i \Gamma_{g\phi}^\lambda 
\left[\bar{\lambda}, \bar{g}, \Phi\right] &=& \int D\psi \int D\bar{c}^\mu Dc^\mu D \bar{d}^\mu Dd^\mu\nonumber \\
& &\exp i \int dx \sqrt{-\bar{g}} \Big\{ \frac{1}{\kappa^2} \Big[
    -\bar{g}^{\mu\nu} R_{\mu\nu}(\bar{g}) + \bar{\lambda}^{\mu\nu}
    \Big( R_{\mu\nu} (\bar{g}) 
- \frac{1}{2}\bar{g}_{\mu\nu} R(\bar{g})\Big)\Big] 
\nonumber \\ &+&
\Big[ \bar{c}^\mu \left( d_{\mu ; \bar{\nu}}^{\;\;\;\;\;;\bar{\nu}} - d_{\nu ; \;\; ;\bar{\mu}}^{\;\;\;\bar{\nu}}\right)
+ \bar{d}^\mu \left( c_{\mu ; \bar{\nu}}^{\;\;\;\;;\bar{\nu}} - c_{\nu ; \;\;\; ;\bar{\mu}}^{\;\;\;\bar{\nu}} \right) \Big]\nonumber \\
&+& \Big[ - \frac{1}{2} \bar{g}^{\mu\nu} \partial_\mu (\Phi + \psi) \partial_\nu (\Phi + \psi) - \frac{m^2}{2}(\Phi+\psi)^2
 - \frac{G}{4!} (\Phi + \psi)^4 \Big]\Big\}\nonumber \\
& &\det{}^{-1} \Big\{ \frac{\delta^2}{\delta h_{\pi\tau} \delta h_{\gamma\delta}} \Big[ -g^{\mu\nu} R_{\mu\nu} (g) - \frac{1}{2\alpha} \left( h_{\mu\alpha ;}^{\;\;\;\;\bar{\alpha}} - k h_{\alpha \;\;\;;\bar{\mu}}^{\;\;\;\alpha} \right)\nonumber \\
& &\left( h^{\mu \;\;\;\;\bar{\beta}}_{\;\;\beta ;} - k h_{\beta \;\;\; ;}^{\;\;\;\beta\;\;\;\;\bar{\mu}} \right)\Big]_{h=0} \Big\}.
\end{eqnarray}
We find from Eq. (\ref{eq87}) that a perturbative expansion of $\Gamma_{g\phi}^\lambda [\bar{g}, \bar{\lambda}, \Phi]$ leads to a structure much like that which follows from Eq. (\ref{eq67}) for YM theory.  We find that the following diagrams contribute to $\Gamma_{g\phi}^\lambda$:
\begin{enumerate}
\item all tree diagrams with background metric $\bar{g}_{\mu\nu}$
and background scalar $\Phi$.
\item twice all one-loop diagrams that follow from the EH action alone, but no diagrams beyond one-loop order.
\item all loop diagrams for the scalar $\Psi$ in the presence of a background metric $\bar{g}_{\mu\nu}$
and background scalar $\Phi$.
\end{enumerate}


We now consider the divergences coming from radiative corrections computed using Eq.~\eqref{eq87}. First of all, the divergences arising from the one-loop diagrams resulting from the EH action alone are \cite{1} 
\begin{equation}\label{eq:5.21}
    \frac{2}{\epsilon} \sqrt{- \bar{g}} \left ( \frac{1}{120} R( \bar{g} ) + \frac{7}{20} R_{\mu \nu} ( \bar{g} ) R^{\mu \nu} ( \bar{g} ) \right ).
\end{equation}
The factor of two arises due to the use of the LM field in Eq.~\eqref{eq78}. From Eq.~\eqref{eq:5.21} and Eq.~(5.22) appearing in ref. \cite{1}, we see that all one-loop divergences involving $ \bar{g}_{\mu \nu} $ resulting from Eq.~\eqref{eq87} are either of the form $ \sqrt{- \bar{g} }F_{\mu \nu \lambda \sigma} ( \bar{g}, \Phi) R^{\lambda \sigma} ( \bar{g} )  $ or $ \sqrt{- \bar{g}} \bar{g}^{\mu \nu} ( \partial_{\mu} \Phi )( \partial_{\nu} \Phi )$. The former divergences can be absorbed into $ \bar{\lambda}^{\mu \nu} $ in Eq.~\eqref{eq87}, the latter by a field renormalization of $ \Phi $. The divergences involving no external metric field can be eliminated by renormalizing $m^2$, $G$ and $ \Phi $.
Higher loop diagrams with contributions coming from the propagation of the scalar field will be proportional to $ G^N$. These diagrams may involve external fields $ \bar{g}_{\mu \nu} $. It should be possible to remove the divergences in such higher-loop diagrams by renormalizing $ m^{2}$, $G$ and $ \Phi $, as those diagrams involve a propagator arising from $ - \sqrt{- \bar{g}} \left ( \bar{g}^{\mu \nu} \partial_{\mu} \Phi \partial_{\nu} \Phi - m^{2} \Phi^{2} \right ) $ and a vertex following from $ - G \sqrt{- \bar{g}} \Phi^{4} $, with the metric not altering the short distance behaviour of divergent diagrams involving $ \Phi $.


If we set the field $\bar\lambda^{\mu\nu}$ equal to zero in Eq.~\eqref{eq87}, we see that all tree-level diagrams are now simply given by the exponential 
\begin{equation}\label{eq:5.22}
  \exp i \int \mathop{dx} \sqrt{- \bar{g}} \left [
    -\frac{1}{\kappa^2} \bar{g}^{\mu \nu} R_{\mu \nu} ( \bar{g} ) - \frac{1}{2} \bar{g}^{\mu \nu} ( \partial_{\mu} \Phi \partial_{\nu} \Phi ) - \frac{m^{2}}{2} \Phi^{2} - \frac{G}{4!} \Phi^{4} \right ] .
\end{equation}
%
We thus see that we have recovered what is expected classically; the background matter field $\Phi$ coupled to the background metric field $ \bar{g}_{\mu \nu} $. Although the classical equation of motion of Eq.~\eqref{A.12} need not be satisfied by $\bar{g}_{\mu \nu}$ and $\Phi$, if we were to examine the classical equation of motion for $\bar{g}_{\mu \nu}$  and $\Phi$ that follow form Eq.~\eqref{eq:5.22},
then $\bar\lambda^{\mu\nu}$ would couple to the tensor ${\bar T}_{\mu \nu}$ of Eq.~\eqref{eq77} formed from $\bar{g}_{\mu \nu}$ and $\Phi$. Since the total background tensor $\bar T_{\mu\nu}$ is covariantly conserved, the
coupling ${\bar\lambda}^{\mu\nu} {\bar T}_{\mu\nu}$
preserves the gauge invariance of the theory. As a result, we obtain Einstein's equations of the gravitational field in a way consistent with the requirements of general relativity.


\section{Discussion}

We have shown that by use of a LM field all radiative corrections beyond one loop order can be eliminated.  Although this necessitates using a renormalization procedure that is inconsistent with unitarity in the case of scalar fields, it is possible to use LM fields to restrict radiative corrections to the YM and EH actions to one-loop order and obtain results that, after using renormalization to remove divergences, are consistent with unitarity.  Furthermore, a scalar field without a LM field can  be coupled in a gauge invariant way to a vector or metric gauge field which has an associated LM field. We anticipated that both vector and spinor matter fields can be incorporated in this manner as well. 

For a pure YM gauge theory, having a LM field results in it being possible to compute the renormalization group functions exactly.  Though this is interesting, one can in fact compute with a YM action without a LM field, as it is both unitary and renormalizable to all orders in the loop expansion.

However, radiative corrections to the EH action alone beyond one-loop order lead to divergences that cannot be removed through renormalization.  It is of interest then to see that a LM field can be used to eliminate those diagrams that result in these higher-loop divergences.  For some time, quantized matter fields have been considered propagating on a curved background whose dynamics is determined by the EH action, possibly supplemented by one-loop corrections \cite{24}; it may even be possible to resolve the Hawking information paradox for black holes using this approach \cite{25}.  We have shown that when one uses a LM field, it is unnecessary to invoke additional fields to cancel higher-loop divergences, as is done for example in supergravity.


At finite temperature, it turns out that both the tree-level as well as the one-loop effects are twice what would come from the EH action alone, which results in the pressure due to the gravitational radiation being doubled
\cite{brandt2021thermal} (to appear in Canadian Journal of Physics). This could possibly be 
a prediction that may be tested experimentally in the foreseeable future.

\begin{acknowledgments}
{Discussions with Roger Macleod were quite helpful.
F. T. B., J. F. and S. M.-F thank CNPq (Brazil) for financial
support. S. M.-F. thanks CAPES (Brazil) for partial financial support.
This study was financed in part by the Coordena\c{c}\~{a}o de Aperfei\c{c}oamento 
de Pessoal de N\'{\i}vel Superior - Brasil (CAPES) - Finance Code 001. This work comes as an aftermath of an original 
project developed with the support of FAPESP (Brazil),  grant number 2018/01073-5.}
\end{acknowledgments}

\appendix

\section{Derivation of the Gauge Symmetries}\label{appA}

In this appendix we show how the Dirac constraint formalism \cite{32}
can be applied to the action of Eq. (\ref{eq28}) to derive the gauge
invariances of Eqs. (\ref{eq16}, \ref{eq29}, \ref{eq30}).

With the metric $\eta_{\mu\nu} = \rm{diag} (-, +, +, +)$, the Lagrangian appearing in Eq. (\ref{eq28}) can be written
\begin{equation}\label{B.1}
\mathcal{L} = \frac{1}{2} f_{0i} f_{0i} - \frac{1}{4} f_{ij} f_{ij} + \lambda_i^a D_j^{ab} f_{ij}^b + \lambda_0^a D_i^{ab} f_{0i}^b + f_{0i}^a D^{ab}_0 \lambda_i^b.
\end{equation}
The canonical momenta are given by
\begin{equation}\label{B.2}
\pi^a =\frac{\partial\mathcal{L}}{\partial (\partial_0 A_0^a)} = 0
\end{equation}
\begin{equation}\label{B.3}
\sigma^a =\frac{\partial\mathcal{L}}{\partial (\partial_0 \lambda_0^a)} = 0
\end{equation}
\begin{equation}\label{B.4}
\pi_i^a = \frac{\partial \mathcal{L}}{\partial(\partial_0 A_i^a)} = f_{0i}^a - D_i^{ab} \left(A_0^b + \lambda_0^b\right) + D_0^{ab} \lambda_i^b
\end{equation}
\begin{equation}\label{B.5}
\sigma_i^a = \frac{\partial \mathcal{L}}{\partial(\partial_0 \lambda_i^a)} = f_{0i}^a .
\end{equation}
As a result we find that
\begin{equation}\label{B.6}
\pi_i^a = \sigma_i^a - D_i^{ab} \lambda_0^b + D_0^{ab} \lambda_i^b.
\end{equation}
The canonical Hamiltonian is given by
\begin{eqnarray}\label{B.7}
\mathcal{H}_c &=& \pi^a \partial_0 A_0^a + \pi_i^a \partial_0 A_i^a + \sigma^a \partial_0\lambda_0^a + \sigma_i^a \partial_0 \lambda_i^a - \mathcal{L} \nonumber\\
&=& \pi^a_i \sigma^a_i - \frac{1}{2} \sigma_i^a \sigma_i^a + \frac{1}{4} f_{ij}^a f_{ij}^a - 
\left(D_i^{ab} \lambda_j^b \right)\left(f_{ij}^a\right)\nonumber \\
&-& \lambda_0^a D_i^{ab} \sigma_i^b - A_0^a \left( D_i^{ab} \pi_i^b + gf^{abc} \lambda_i^b \sigma_i^c\right).
\end{eqnarray}
From Eqs. (\ref{B.2}, \ref{B.3}) we immediately have the primary constraints \cite{32}
\begin{equation}\label{B.8}
\phi_I^a = \pi^a
\end{equation}
\begin{equation}\label{B.9}
\phi_{II}^a = \sigma^a  .
\end{equation}
In order for these primary constraints to be constant in time, their Poisson bracket (PB) with $\mathcal{H}_c$ must either vanish or else new (secondary) constraints are present.  Using the PB
\begin{equation}\label{B.10}
\left\lbrace A_\mu^a (\mathbf{x},t), \pi_\nu^b(\mathbf{y},t)\right\rbrace = \delta^{ab} \eta_{\mu\nu} = \left\lbrace \lambda_\mu^a (\mathbf{x}, t), \sigma_\nu^b (\mathbf{y},t)\right\rbrace
\end{equation}
this leads to the secondary constraints
\begin{equation}\label{B.11}
\Phi_I^a = D^{ab}_i \pi_i^b + gf^{abc} \lambda_i^b \sigma_i^c
\end{equation}
\begin{equation}\label{B.12}
\Phi_{II}^a = D^{ab}_i \sigma_i^b .
\end{equation}
No tertiary constraints need to be introduced.  These constraints are all first class, as their PB with each other either vanish, or else vanish on the ``constraint surface'' as
\begin{equation}\label{B.13}
\left\lbrace \Phi_I^a, \Phi_I^b\right\rbrace = gf^{abc} \Phi_I^c
\end{equation}
\begin{equation}\label{B.14}
\left\lbrace \Phi_I^a, \Phi_{II}^b\right\rbrace = gf^{abc} \Phi_{II}^c   .
\end{equation}

The ``total Hamiltonian'' $\mathcal{H}_T$ and ``extended Hamiltonian'' $\mathcal{H}_E$ are now defined
\begin{equation}\label{B.15}
\mathcal{H}_T = \mathcal{H}_c + x_i^a \phi_I^a + x_{II}^a \phi_{II}^a
\end{equation}
\begin{equation}\label{B.16}
\mathcal{H}_E = \mathcal{H}_T + X_I^a \Phi_I^a + X_{II}^a \Phi_{II}^a
\end{equation}
with $x_I^a$, $x_{II}^a$, $X_I^a$, $X_{II}^a$ being treated as dynamical variables whose equation of motion simply ensures that the constraints are satisfied. If we consider the Hamilton equations of motion that follow from the actions
\begin{equation}\label{B.17}
S_T = \int d^3x \left[ \pi^a \partial_0 A_0^a + \sigma^a \partial_0 \lambda_0^a + \pi^a_i \partial_0 A_i^a + \sigma^a_i \partial_0 \lambda_i^a - \mathcal{H}_T \right]
\end{equation}
\begin{equation}\label{B.18}
S_E =S_T - \int d^3x \left( X_I^a \Phi_I^a + X_{II}^a \Phi_{II}^a \right)
\end{equation}
then the Hamilton equations of motion that follow from Eq. (\ref{B.17}) have the same dynamical content as the Lagrangian equations of motion that follow from $S_{2YM}^\lambda$ in Eq. (\ref{eq28}).

In general, if we have first class constraints $\gamma_{a_i}$ in the i$^{th}$ generation, then we will show that a generator $G$ of the form
\begin{equation}\label{B.19}
    G = \int \mathop{d^3x} \rho_{a_i} \gamma_{a_i} 
\end{equation}
will lead to local transformations of dynamical variables $\phi_A$ and their conjugate momenta $\pi_A$ of the form
\begin{equation}\label{B.20}
\delta F = \left\lbrace F,G \right\rbrace
\end{equation}
that leave $S_T$ invariant, thereby providing the gauge invariances of the action.  Following the HTZ approach \cite{33},
\begin{eqnarray}\label{B.21}
\delta S_E &=& \delta \int d^3x dt \left( \pi_A \partial_0 \phi_A - \mathcal{H}_c - U_{a_i} \gamma_{a_i}\right)\nonumber \\
&=& \int d^3x dt \Bigg[ \left\lbrace \pi_a, G\right\rbrace \partial_0 \phi_A - \left\lbrace \pi_a, G\right\rbrace 
\partial_0 \pi_A 
-  \left\lbrace \mathcal{H}_c, G\right\rbrace - U_{a_i} \left\lbrace
                                  \gamma_{a_i}, G\right\rbrace -
                                  \delta U_{a_i}\gamma_{a_i}\Bigg] 
\end{eqnarray}
\begin{equation}\label{B.22}
= \int d^3x dt \Bigg[ - \frac{\partial G}{\partial\phi_A} \partial_0
  \phi_A - \frac{\partial G}{\partial\pi_A} \partial_0\pi_A -
  \left\lbrace \mathcal{H}_c, G\right\rbrace  
- U_{a_i} \left\lbrace \gamma_{a_i}, G\right\rbrace  - \delta U_{a_i} \gamma_{a_i}\Bigg]. 
\end{equation}
If now
\begin{equation}\label{B.23}
\left\lbrace U_{a_i}, G \right\rbrace = 0,
\end{equation}
and since
\begin{equation}\label{B.24}
\frac{dG}{dt} = \int d^3x \left[ \dot{\rho}_{a_i} \gamma_{a_i} + \frac{\partial G}{\partial \phi_A} \partial_0 \phi_A + 
\frac{\partial G}{\partial \pi_A} \partial_0 \pi_A\right]
\end{equation}
then
\begin{equation}\label{B.25}
\delta S_E = \int d^3x \left[ \dot{\rho}_{a_i} \gamma_{a_i} + \left\lbrace G, \mathcal{H}_E \right\rbrace - \delta U_{a_i} \gamma_{\sigma_i}\right] .
\end{equation}
If now we specialize to the case where $U_{a_i} = 0 = \delta U_{a_i}$ for $i > 1$, then we obtain
\begin{equation}\label{B.26}
\delta S_T = \int d^3 x \left[ \dot{\rho}_{a_i} \gamma_{a_i} +  \left\lbrace G, \mathcal{H}_T \right\rbrace - \delta U_{a_1} \gamma_{a_1}
\right]. 
\end{equation}
If we use Eq. (\ref{B.26}) to solve for $\rho_{a_i}$ so that $\delta S_T = 0$, we have the generator $G$ that leaves the action invariant.

With $\mathcal{H}_c$ of Eq. (\ref{B.7}), $G$ is of the form
\begin{equation}\label{B.27}
G = \rho_I^a \phi_I^a + \rho_{II}^a \phi_{II}^a + R_I^a \Phi_I^a + R_{II}^a \Phi_{II}^a .
\end{equation}
From Eq. (\ref{eq26}), we see that $\rho_I^a$ and $\rho_{II}^a$ can be found in terms of $R_I^a$ and $R_{II}^a$; the PB of Eqs. (\ref{B.13}, \ref{B.14}) show that
\begin{equation}\label{B.28}
\rho_I^a = -D_0^{ab} R_I^b
\end{equation}
\begin{equation}\label{B.29}
\rho_{II}^a = -D_0^{ab} R_{II}^b - gf^{abc} \lambda_0^b R_I^c
\end{equation}
leading to
\begin{equation}\label{B.30}
G = R_I^a \left[ D_\mu^{ab} \pi^{b\nu} + gf^{abc} \lambda_\mu^b \sigma^{c\mu}\right]+ R_{II}^a \left[ D_\mu^{ab} \sigma^{b\mu}\right],
\end{equation}
which gives the gauge transformations of Eqs. (\ref{eq16}, \ref{eq19}, \ref{eq30}) provided
\begin{equation}\label{B.31}
\xi^a = R_I^a
\end{equation}
\begin{equation}\label{B.32}
\zeta^a = R_{II}^a.
\end{equation}
Using this same approach, it should be possible to derive the
transformations of Eqs. (\ref{eq70}, \ref{eq81}, \ref{eq82}) for the action of Eq. (\ref{eq78}).

\section{Unitarity with multiplier fields}\label{appB}


The introduction of a LM field to ensure that the classical equation of motion is satisfied leads to three possible problems when the theory is quantized. First of all, one must show that the energy of the system is bounded below. Secondly, it is necessary to demonstrate that states with negative norm \cite{47,48} do not contribute to physical processes. Finally, proving that the contributions of unphysical polarizations of gauge fields and the contributions of ghost fields all cancel is necessary. In this appendix we will deal with each of these issues.

To demonstrate how the energy is bounded below and that negative norm states do not contribute when a LM field is introduced, we consider a very simple model whose action is 
\begin{equation}\label{eq:c1}
    S_{ \text{cl}} = \int \mathop{dx} \left[\frac{1}{2} ( \partial_{\mu} A)^{2} - \frac{m^{2}}{2} A^{2} - \frac{G}{3!} A^{3} + B \left( - \partial^{2} A - m^{2} A - \frac{G}{2!} A^{2}\right)\right]
\end{equation}
(We ignore the need of having two distinct masses and coupling an account of the way renormalization is effected, and the unboundedness of the cubic potential.).
The treatment of the path integral of Eq.~\eqref{eq5} can be repeated here in order to show that all radiative effects beyond one-loop order do not arise.

One can also use canonical quantization of get this result. The momenta conjugate to $A$ and $B$ are $\pi  = \partial_{0} (A+B)$ and  $\sigma = \partial_{0} A$ respectively. If quantization is effected by imposing the usual commutation relations on $A$ and $ \pi $ and on $ B$ and $ \sigma $, we find that Fourier transforms of $A$ and $B$ satisfy
\begin{subequations} \label{eq:c2}
\begin{align}\label{eq:c2a}
                \left [ a ( \mathbf{k} ), a^{\dagger}  ( \mathbf{k'} ) \right ] &= 0,
                \\ \label{eq:c2b}
                \left [ a ( \mathbf{k} ), b^{\dagger}  ( \mathbf{k'} ) \right ] &= \delta ( \mathbf{k} - \mathbf{k'} ),
                \\ \label{eq:c2c}
                \left [ b( \mathbf{k} ), b^{\dagger}  ( \mathbf{k'} ) \right ] &= -\delta ( \mathbf{k} - \mathbf{k'} ).
            \end{align}
        \end{subequations}
            Eq.~\eqref{eq:c2} is consistent with the Feynman propagators in Fig. 1.

            With the action of Eq.~\eqref{eq:c1}, the $S$-matrix is given by (see, for example, \cite{46})
with 
        \begin{equation}\label{eq:c3}
            S = T \exp \left [ - i \int_{- \infty}^{\infty} \mathop{dt} H^{(I)}  \right ],
        \end{equation}
        where
        \begin{equation}\label{eq:c4}
            H^{(I)} = G\int \mathop{d^{} \mathbf{x} } \left ( \frac{1}{3!} A^{3} + \frac{1}{2!} B A^{2} \right ). 
        \end{equation}
        Upon expanding the exponential of Eq.~\eqref{eq:c4} in powers of $G$ and using Wick's theorem to expand the $T$ product in Eq.~\eqref{eq:c3} in terms of normal ordered products and Feynman propagators (as in ref. \cite{46}) one can see that if we examine Green's functions with only external fields A, only one-loop diagrams can contribute, and that those only involve the propagator $ \left \langle AB \right \rangle $. For example, the two-point function only receives the contribution of Fig. 4.

\begin{figure}[ht]
\includegraphics[scale=0.6]{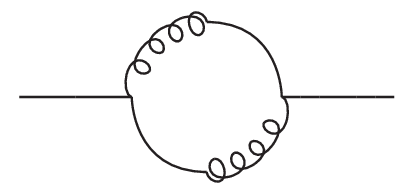} 
\caption{One-loop contribution to $ \left \langle AA \right \rangle $. Solid lines represent $A$, springy lines $B$.}
\end{figure}

If now we treat $ a^{\dagger} ( \mathbf{k} )$ and $ b^{\dagger} ( \mathbf{k} )$ as creation operators, then we have states \cite{47}  
\begin{subequations}\label{eq:c5} 
            \begin{align}\label{eq:c5a}
                \sqrt{n_{k}^{a}!} \left | n_{k}^{a} \right \rangle =[a^{\dagger} ( \mathbf{k} )]^{n^{a}_{k}} \left| 0 \right \rangle  ,
                \\ \label{eq:c5b}
                \sqrt{n_{k}^{b}!} \left | n_{k}^{b} \right \rangle =[b^{\dagger} ( \mathbf{k} )]^{n^{b}_{k}} \left| 0  \right \rangle  ,
            \end{align}
        \end{subequations}
        resulting in 
        \begin{equation}\label{eq:c6}
            \left\langle n_{k}^{a}  \middle | n_{k'}^{a}  \right\rangle 
            = \delta_{n_{k}^{a} n_{k'}^{a}} \delta ( \mathbf{k} - \mathbf{k'} ) 
        \end{equation}
        and
        \begin{equation}\label{eq:c7}
            \left\langle n_{k}^{b}  \middle | n_{k'}^{b}  \right\rangle 
            = (-1)^{n_{k}^{b}} \delta_{n_{k}^{b} n_{k'}^{b}} \delta ( \mathbf{k} - \mathbf{k'} ) 
            . 
        \end{equation}
        Furthermore, upon expressing $H_0$, the free Hamiltonian that follow from Eq.~\eqref{eq:c1} in terms of creation and annihilation operators, it follows that 
        \begin{equation}\label{eq:c8}
            H_{0} 
                        = \frac{1}{2} \int \mathop{d \mathbf{k}} \omega_{k} \left [ ( a(\mathbf{k}) + b(\mathbf{k})) ( a^\dagger(\mathbf{k}) + b^\dagger(\mathbf{k}) ) + ( a^\dagger(\mathbf{k}) + b^\dagger(\mathbf{k}) ) ( a(\mathbf{k}) + b(\mathbf{k})) - ( b ( \mathbf{k} )b^\dagger(\mathbf{k}) + b^\dagger(\mathbf{k}) b(\mathbf{k})) \right ],
        \end{equation}
        where $ \omega_{k} \equiv \sqrt{ \mathbf{k}^{2} + m^{2}} $. 
        Together, Eqs.~\eqref{eq:c2} and \eqref{eq:c8} show that
\begin{subequations}\label{eq:c9} 
            \begin{align}\label{eq:c9a}
                \left[ H_{0} , a^{\dagger} ( \mathbf{k} ) \right] &=  a^{\dagger} ( \mathbf{k} ),
                \\ \label{eq:c9b}
                \left[ H_{0} , b^{\dagger} ( \mathbf{k} ) \right] &=  b^{\dagger} ( \mathbf{k} ).
            \end{align}
        \end{subequations}
and so
        \begin{equation}\label{eq:c10}
            H_0 \left | n^{a}_{k} , n^{b}_{k'} \right \rangle = \left [ \omega_{k} \left( n^{a}_{k} + \frac{1}{2}\right) + \omega_{k'} \left( n_{k'}^{b} + \frac{1}{2}\right) \right ]  \left | n^{a}_{k} , n^{b}_{k'} \right \rangle.
        \end{equation}
        By Eq.~\eqref{eq:c10} we see that the energy spectrum of free particle states, when there is a LM field, is positive definite.

        From Eq.~\eqref{eq:c7}, it follows that if $ n_{k}^{b} $ is odd, there is a negative norm state. However, it is clear from the Feynman rules that no external state will have contributions that involve the states $ | n_{k}^{b} \rangle $ with $ n_{k}^{b} $ odd.

        This can all be seen more clearly if in Eq.~\eqref{eq:c1} we were to replace $A$ with $C - B$, so that 
        \begin{equation}\label{eq:c11}
            S_{ \text{cl}} = \int \mathop{dx} \left[\frac{1}{2} \left(( \partial_{\mu} C)^{2} - m^{2}C^{2}\right)
-\frac{1}{2} \left(( \partial_{\mu} B)^{2} - m^{2}B^{2}\right)
-  G \left( \frac{C^{3}}{6} - \frac{CB^{2}}{2}+\frac{B^3}{3}\right)\right].
        \end{equation}
        In this form of $ \mathcal{L} $, if we were to take the momenta conjugate to $C$ and $B$ to be $ \rho = \partial_{0} C $ and $ \sigma =- \partial_{0} B$, then the usual commutation rules 
        \begin{equation}\label{eq:c12}
             \left[ C ( \mathbf{x}, t), \rho ( \mathbf{y}, t) \right] 
            =
\left [ B( \mathbf{x} , t ), \sigma ( \mathbf{y} ,t ) \right ] =  i \delta ( \mathbf{x} - \mathbf{y})
        \end{equation}
        lead to \cite{47}
\begin{subequations}\label{eq:c13} 
            \begin{align}\label{eq:c13a}
                \left [ c ( \mathbf{k} ), c^{\dagger}  ( \mathbf{k'} ) \right ] &= \delta ( \mathbf{k} - \mathbf{k'} ),
                \\ \label{eq:c13b}
                \left [ b( \mathbf{k} ), b^{\dagger}  ( \mathbf{k'} ) \right ] &= -\delta ( \mathbf{k} - \mathbf{k'} ),
            \end{align}
        \end{subequations}
        where we have the Fourier transform of $ C ( \mathbf{x} ,t )$ and $ B ( \mathbf{x} ,t) $ in Eq.~\eqref{eq:c13}. Treating $ c^{\dagger} $ and $ b^{\dagger} $ to be creation operators as in Eq.~\eqref{eq:c5}, we recover Eqs.~\eqref{eq:c6} and \eqref{eq:c7} with $ n^{a}_{k} $ replaced by $ n_{k}^{c} $. The free Hamiltonian of Eq.~\eqref{eq:c8} now 
        \begin{equation}\label{eq:c14}
            H_0= \frac{1}{2} \int \mathop{d \mathbf{k}} \omega_{k} 
            \left [ ( c(\mathbf{k}) c^\dagger(\mathbf{k}) + c^\dagger(\mathbf{k}) c(\mathbf{k})) - ( b(\mathbf{k}) b^\dagger(\mathbf{k}) + b^\dagger(\mathbf{k}) b(\mathbf{k})) \right ] .
        \end{equation}
        We can now recover Eqs.~\eqref{eq:c9} and \eqref{eq:c10} with $ a^{\dagger} $ and $ n_{k}^{a} $ replaced by $c^{\dagger} ( \mathbf{k} )$ and $ n_{k}^{c} $ respectively, so we can again conclude that the energy spectrum is bounded below.

        To see that only an even number of states of the form $ | n_{k}^{b} \rangle $ contribute to any process, we use that path integral of Eq.~\eqref{A.23} to quantize the system of Eq.~\eqref{eq:c11}. If $ \bar{C} $ is the background field for $C$, 
        \begin{equation}\label{eq:c15}
            C = \bar{C} + \gamma , \quad B = \beta
        \end{equation}
        we then have the generating functional 
        \begin{equation}\label{eq:c16}
            \begin{split}
                e^{i \Gamma [ \bar{C} ]} &= e^{i \int \mathop{dx} \left[\frac{1}{2} ((\partial_{\mu} \bar{C} )^{2} - m^{2} \bar{C}^{2} ) - \frac{G}{3!} \bar{C}^{3}\right]} \\
                                         & \times \int \mathop{D \gamma } \mathop{D \beta} \exp i \int \mathop{d^{}x} \left[
\frac{1}{2} ((\partial_{\mu} \gamma )^{2} - m^{2} \gamma^{2} ) 
-
\frac{1}{2} ((\partial_{\mu} \beta )^{2} - m^{2} \beta^{2} ) 
-G \frac{ \bar{C} }{2} ( \gamma^{2} - \beta^{2} )
            \right].
        \end{split}
        \end{equation}
        In Eq.~\eqref{eq:c16} we have dropped all terms linear in the quantum fields as well as those that can only contribute to Feynman graph with more than one loop. (In ref. \cite{21a} explicit calculations demonstrated that such higher loop graphs sum to zero in this model.) It is clear from the fact that since the action appearing in Eq.~\eqref{eq:c16} is even in the quantum field $ \beta $, all physical processes in this model must involve an even number of excitations of the form $ | n_{k}^{b} \rangle $. More explicitly, since the only vertices are of form $- G \bar{C} \gamma^{2} /2$ and $+G \bar{C} \beta^{2} /2$, and the propagators for $ \gamma $ and $ \beta $ are $ i/( k^{2} -m^{2} )$ and $ - i ( k^{2} -m^{2} )$ respectively we see that the only possible Feynman graphs involve external fields $ \bar{C} $ and either a loop of fields $ \gamma $ or $ \beta $, with these two loops summing to give twice the contribution of the one-loop graphs that arise if there were no LM field $B$ in Eq.~\eqref{eq:c1}. This is consistent with Eq.~\eqref{A.26}.

        The arguments that we have used to establish that quantizing the action of Eq.~\eqref{eq:c1} leads to a bounded energy spectrum and to absence of negative norm states can be applied to more involved models, such as the gauge theories defined by the YM and EH actions. These too will have a bounded energy spectrum and will not involve negative norm states.

        In gauge theories an extra complication occur, for in order to establish unitarity, it is necessary to show that the contributions of ghost fields and non-physical polarizations of gauge fields must all cancel. We will now establish how this happens when the usual gauge action is supplemented by a LM field.


For simplicity, we consider first the theory in the standard second order formulation,
which is described by the effective Lagrangian \cite{4}
\begin{eqnarray}\label{c1}
    {\cal L}_{\text{eff}} &=& -\frac 1 4 f_{\mu\nu}^a f^{\mu\nu\,a} -  \lambda_\mu^a D^{ab}_\nu(A) f^{b\, \mu\nu}
-N^a\partial\cdot(A^a+\lambda^a)+\frac\alpha 2 N^a N^a -
                   L^a\partial\cdot A^a + \alpha N^a L^a \nonumber \\ 
&-&{\bar c}^a\partial \cdot D^{ab}(A) d^b - {\bar d}^a\partial \cdot D^{ab}(A) c^b
- {\bar c}^a\partial \cdot D^{ab}(A+\lambda) c^b ,
\end{eqnarray}
where $f_{\mu\nu}^a$ and $D_\nu^{ab}$ are given by Eqs. \eqref{eq15}
and \eqref{eq17}, and $\lambda^a_\mu$ is the Lagrange multiplier field. 
The third, fourth, fifth and sixth terms define the gauge fixing part of the Lagrangian
(using the auxiliary Nakanish-Lautrup fields $N^a$ and $L^a$)
while the last three terms fix the ghost sector of the Lagrangian, which is  obtained by
using the Faddeev-Popov procedure.
The complete Lagrangian \eqref{c1} is invariant under the BRST transformations
\begin{subequations}\label{c2}
\be\label{c2a}
Q A^a_\mu = \delta A_\mu^a \zeta = D_\mu^{ab}(A) c^b \zeta,
\ee
\be\label{c2b}
Q \lambda_\mu^a =
\delta \lambda_\mu^a \zeta = \left[D_\mu^{ab}(A) d^b + g f^{abc} \lambda_\mu^b c^c\right]
\zeta , 
\ee
\be\label{c2c}
Q N^a = Q L^a = 0 ,
\ee
\end{subequations}
where $\zeta$  is an infinitesimal Grassmann constant, and
\begin{subequations}\label{c3}
\be\label{c3c}
Q {\bar c}^a = \delta {\bar c}^a \zeta = \partial\cdot A^a\zeta,
\ee
\be\label{c3d}
Q {\bar d}^a = \delta {\bar d}^a \zeta = \partial\cdot \lambda^a\zeta,
\ee
\be\label{c3e}
Q c^a = \delta {c}^a \zeta = \frac g 2 f^{abc} c^b c^c\zeta,
\ee
\be\label{c3f}
Q d^a = \delta {d}^a \zeta =  g  f^{abc} c^b d^c\zeta .
\ee
\end{subequations}
It may be verified that the above BRST transformations are nilpotent \cite{21a}. 

In the Feynman gauge, $\alpha=1$, the quadratic part of the Lagrangian is 
\be\label{c4}
{\cal L}^{(2)} =
\frac 1 2 \left(\begin{array}{cc} A_\mu^a, & \lambda_\mu^a \end{array}\right) 
\left(\begin{array}{cc} \partial^2 & \partial^2  \\ \partial^2 & 0\end{array}\right) 
\left(\begin{array}{c} A_\mu^a  \\ \lambda_\mu^a \end{array}\right)
-
\left(\begin{array}{cc} {\bar c}^a, & {\bar d}^a \end{array}\right) 
\left(\begin{array}{cc} \partial^2 & \partial^2  \\ \partial^2 & 0\end{array}\right) 
\left(\begin{array}{c} c^a  \\ d^a \end{array}\right) .
\ee
We have that
\be\label{c5}
\left(\begin{array}{cc} \partial^2 & \partial^2  \\ \partial^2 & 0\end{array}\right)^{-1}= 
\left(\begin{array}{cc} 0 & \frac{1}{\partial^2-i\epsilon}  \\ \frac{1}{\partial^2-i\epsilon}   & -\frac{1}{\partial^2-i\epsilon}  \end{array}\right) ,
\ee
where we have made explicit the $i\epsilon$ prescription. 
Thus, we see that at the tree level there are $\langle \lambda\lambda \rangle$
and $\langle {\bar d} d\rangle$ propagators, as well as mixed 
$\langle A\lambda \rangle$, $\langle {\bar c} d \rangle$  and $\langle {\bar d} c \rangle$ propagators, but no $ \left \langle AA \right \rangle$ or $  \left \langle \bar{c} c \right \rangle$ propagators.
This fact, and the absence of any vertices with more than one external
field $\lambda^a_\mu$, $d^a$ or ${\bar d}^a$implies that there are no diagrams beyond 
one loop order \cite{4,21,21a}.

We note here that the $\langle \lambda\lambda \rangle$
propagator occurs with a negative metric. However, it can
be shown \cite{4} that only the first two terms in the ghost sector of the Lagrangian \eqref{c1} do
contribute to the generating functional, so that the multiplier field  $\lambda$  actually decouples
from the ghost sector. This fact, together with the previous features,
ensures that there are no one-loop diagrams which contain the
$\langle \lambda\lambda \rangle$ propagator. This property is quite
relevant for the unitarity of the theory.

Let us now consider, as an example, the amputated two-point function shown in Fig. 5.
\begin{figure}[t]
\includegraphics[scale=0.55]{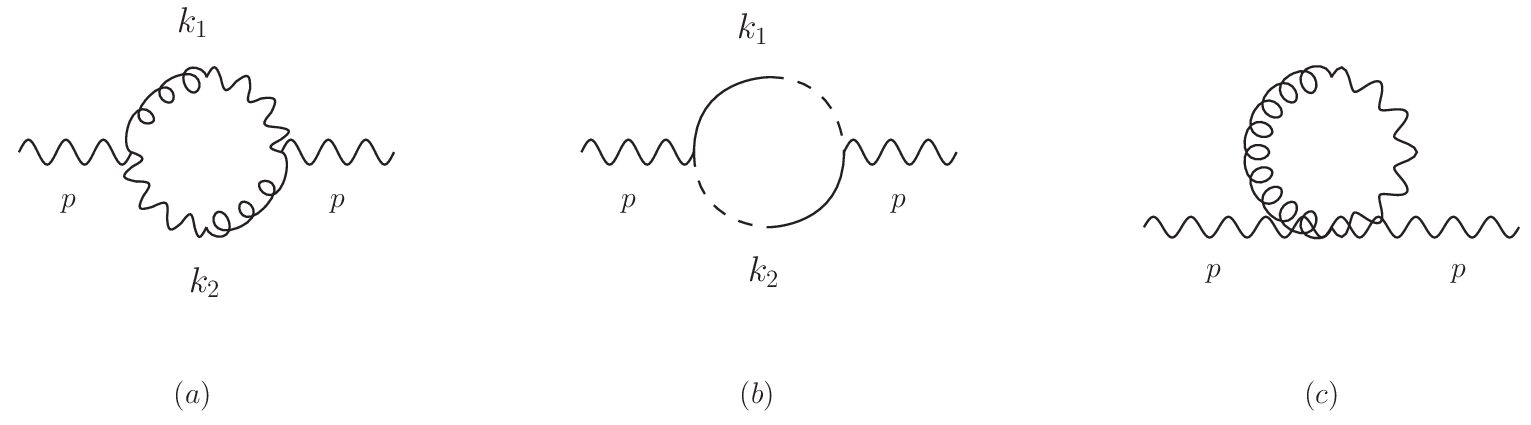} 
\caption{One-loop graphs for the gluon self-energy. Springy lines denote the multiplier field, 
full lines represent the  $d$  ghosts  and dashed lines stand for the $c$ ghosts.}
\end{figure}
The tadpole contribution from Fig. 5c  vanishes when using dimensional regularization.
The other contributions from the above one-loop graphs turn out to be twice those obtained in the pure YM theory.
From unitarity, one would expect that the imaginary part of the self-energy to be related to the T-matrix as
\be\label{c6}
2 \Im\langle p|T|p \rangle = \int\frac{d^3 k_1}{(2\pi)^3}\frac{\theta(k_1^{(0)})}{2 k_1^{(0)}}
\int\frac{d^3 k_2}{(2\pi)^3}\frac{\theta(k_2^{(0)})}{2 k_2^{(0)}}
(2\pi)^4\delta(p-k_1-k_2) \langle p|T|k_1 k_2 \rangle {\langle p|T|k_1
  k_2 \rangle}^\star . 
\ee
Here, the integrations are only over the momenta 
of transverse gauge bosons $A$ and $\lambda$ which are on-shell with positive energies,
as depicted in Fig. 6.
\begin{figure}[b]
\includegraphics[scale=0.55]{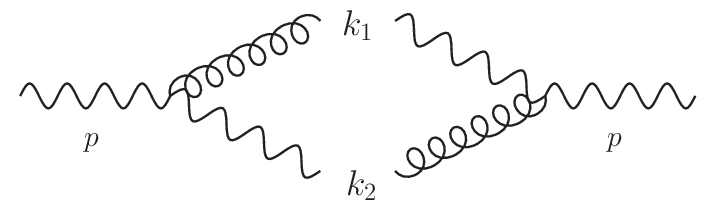} 
\caption{Diagram illustrating the gluon decay process $A\rightarrow 
  A+\lambda$ in the multiplier theory.}
\end{figure}
The above contribution is also twice that obtained in the pure YM
theory, since in this case there would be an extra factor of $1/2$ to account for the identical gluons in the final state.
   
In order to verify the unitarity condition, let us consider the
Cutkosky \cite{Cutkosky:1960sp} cut diagrams 
shown in Fig. 7, which are associated with the graphs shown in Fig. 6.
\begin{figure}[t]
\includegraphics[scale=0.55]{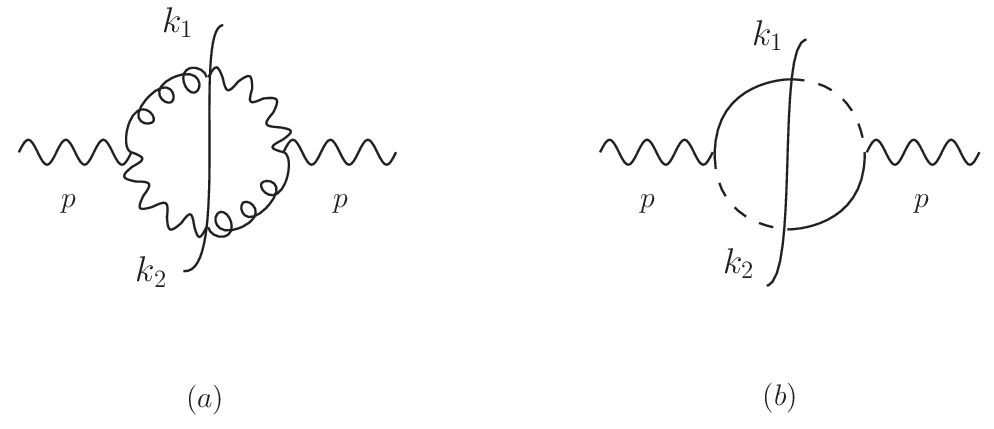} 
\caption{Graphical representation of Cutkosky cut diagrams.}
\end{figure}
Here, cutting a propagator with momenta $k$ yields for an out-going particle a factor like
\be\label{c7}
\mbox{cut propagator}(k^{\mbox{out}}) = 
\Im \frac{g^{\mu\nu}}{k^2+i\epsilon} \theta(k_0) = -i\pi g^{\mu\nu}
\delta(k^2) \theta(k_0) .
\ee
Thus, the propagator is replaced by a mass-shell contribution with positive energy for an
outgoing gauge particle. It is important to note that all four polarization states occur here.
      
To check unitarity, we use the BRST symmetry of our theory as well as
the nilpotency of the  BRST transformation. The BRST operator gives a precise relation between the 
unphysical gauge boson polarization states and the ghost and anti-ghost
degrees of  freedom. This leads to a cancellation of diagrams
involving unphysical longitudinal and timelike gauge bosons with those containing ghost and anti-ghost fields 
\cite {31,Nishijima:1978wq,Nakanishi:1990qm,peskin_scroeder,weinberg:book2005,Srednicki}.
As a result, the sum of the Cutkosky cut diagrams shown in Fig. 7
yield the same unitary result as  that given by Fig. 6.

Let us now consider an example of a a two-loop graph for the gluon self-energy, shown
in Fig. 8a, which vanishes due to the fact  there is no  $\langle A A \rangle$ propagator.
\begin{figure}[h!]
\includegraphics[width=\textwidth]{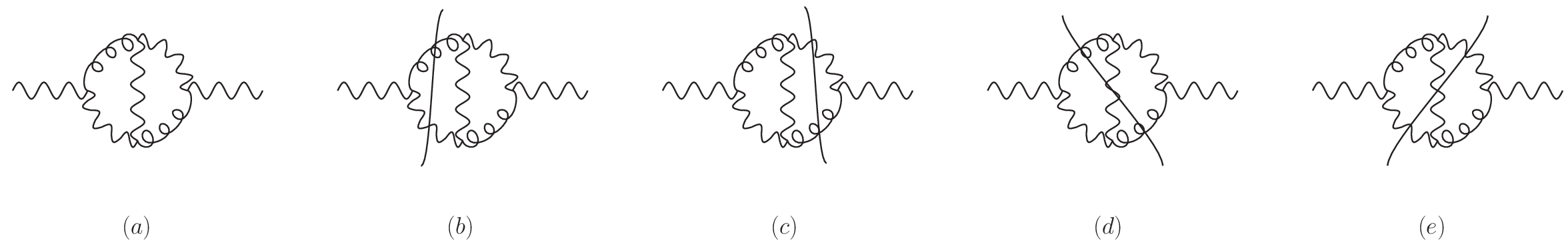} 
\caption{A two-loop self energy graph and its associated cut diagrams.}
\end{figure}
Cutting the graphs in Fig. 8a yields the diagrams shown in
Figs. (8b), (8c), (8d) and (8e). The diagrams (8b) and (8c) vanish due to
the fact that there is no $\langle AA \rangle$ propagator, while 
the diagrams  (8d) and (8e)  vanish since, correspondingly, there are no cut gluon propagators.
Thus, in the Lagrange multiplier theory, the vanishing of higher-order
loops yield self-consistent results which do not violate unitarity. 

For completeness, we have also considered the two-point function in a
scalar $A^3$ theory 
with Lagrange multiplier fields.
We have explicitly verified that the two-point function at one-loop order
satisfies the unitarity condition that the cut diagram is the square
of the modulus of the appropriate tree diagram.
This is discussed in more detail below.

Having shown how the two-point function is consistent with unitarity
through explicit computation, we will consider how unitarity can be
established by using the BRST transformation of Eqs. \eqref{c2} and
\eqref{c3} \cite{30,31}, following the presentation in Ref. \cite{peskin_scroeder}.

The BRST transformations of Eqs. \eqref{c2} and \eqref{c3} satisfy the
condition
\be\label{c8}
Q^2 = 0,
\ee
by which we mean that for any field $\phi_i$,
\be\label{c9}
Q(\delta \phi_i) = 0.
\ee
(One writes the gauge fixing terms in the effective Lagrangian ${\cal
L}_{\text{eff}}$ in terms of the auxiliary Nakanishi-Lautrup fields $N^a$
and $L^a$ in order that Eqs. \eqref{c8} and \eqref{c9} are satisfied.)

Due to ${\cal L}_{\text{eff}}$ being invariant under the BRST
transformations, the operator $Q$ commutes with the Hamiltonian, and
also with the $S$ matrix. We now consider distinct types of states.

First of all we will consider ``physical states'' $|\psi_P\rangle$ which
satisfy 
\be\label{c10}
Q | \psi_P\rangle = 0 .
\ee
We then will consider two types of states 
$|\psi_P\rangle$ satisfying Eq. \eqref{c8}. First, if
$|\psi_{nP}\rangle$ is a ``non-physical state'' that does not satisfy
Eq. \eqref{c10}, then $|\psi_{P^\prime}\rangle$ is such that 
\be\label{c11}
| \psi_{P^\prime} \rangle \equiv Q |  \psi_{nP} \rangle .
\ee
Secondly, there are states $| \psi_{P_0} \rangle$ that satisfy
Eq. \eqref{c10} but cannot be written as in Eq. \eqref{c11}.
It is evident from Eqs. \eqref{c8} and \eqref{c11} that two states 
$| \phi_{P^\prime}\rangle$ and   $|\psi_{P^\prime} \rangle$ have vanishing
inner product
\be\label{c12}
\langle \phi_{P^\prime} | \psi_{P^\prime} \rangle=
\langle \phi_{nP} | Q^2 | \psi_{nP} \rangle = 0 .
\ee

It is evident from Eqs. \eqref{c8} and \eqref{c11} that 
the inner product of two states  $|\psi_{P_0}\rangle$ and $|\psi_{P^\prime}\rangle$
vanishes by Eqs. \eqref{c10} and \eqref{c11}, 
\be\label{c13}
\langle \psi_{P_0} | \psi_{P^\prime} \rangle = 
\langle \psi_{P_0} |Q| \psi_{nP} \rangle =0 .
\ee

In the limit $g=0$, Eqs. \eqref{c2a} and \eqref{c2b} show that
longitudinal polarizations of $A_\mu^a$ and $\lambda_\mu^a$
are converted into ghost fields $c^a$ and $d^a$; these in turn are
annihilated by $Q$ when $g=0$ by Eqs. \eqref{c3c} and \eqref{c3d}.
So also, the ghost fields ${\bar c}^a$ and ${\bar d}^a$, when operated on
by $Q$, become the auxiliary fields $N^a$ and $L^a$ by
Eqs. \eqref{c2a} and \eqref{c2b}; these in turn are the longitudinal
components of $A_\mu^a$ and $\lambda_\mu^a$
as can be seen from the equations of motion for $N^a$ and $L^a$ that
follow from Eq. \eqref{c1}.

Consequently, we see that states with longitudinal polarizations of
$A^a_\mu$ and $\lambda_\mu^a$ are of the type $|\psi_{P^\prime}\rangle$,
the transverse polarizations of $A_\mu^a$ and $\lambda_\mu^a$
are in states of the type $|\psi_{P_0}\rangle$,
the ghosts $c^a$ and $d^a$ are of the type $|\psi_{P^\prime}\rangle$
and
${\bar c}^a$ and ${\bar d}^a$ are of the type $|\psi_{nP}\rangle$, all
at $g=0$.
This can be more precisely worked out by following the arguments in Ref. \cite{31}.

If we initially have a single particle state $|\psi_P\rangle$
satisfying Eq. \eqref{c10}, then since $Q$ commutes with $S$, we also
will have 
\be\label{c14}
Q S | \psi_P\rangle =0,
\ee
and thus $S| \psi_P\rangle$ is either in a state $| \psi_{P_0}\rangle$ 
or $| \psi_{P^\prime}\rangle$.
As a result of this, and the fact that $| \phi_{P^\prime}\rangle$ 
is orthogonal to $| \psi_{P_0}\rangle$  and $| \psi_{P^\prime}\rangle$ 
(by Eqs. \eqref{c12} and \eqref{c13}) we see that in the sum
\be\label{c15}
\langle \phi_{P_0}|S^\dagger S |\psi_{P_0} \rangle=
\sum_{\chi}\langle \phi_{P_0}|S^\dagger|\chi\rangle\langle\chi| S |\psi_{P_0}\rangle
\ee
$|\chi\rangle$ must be a linear combination of the
states $|\chi_{P_0}\rangle$ and $|\chi_{P^\prime}\rangle$. 
However, the states $|\chi_{P^\prime}\rangle$ 
have, by \eqref{c12} and
\eqref{c13}, zero inner product with one another and with $|\chi_{P_0}\rangle$. 
Thus,  only $|\chi_{P_0}\rangle$ will make a non-zero contribution to the
inner products. As a result, {\it only transverse polarizations of
$A_\mu^a$ and $\lambda_\mu^a$ can contribute in the cutting
relations}, with longitudinal and timelike polarizations cancelling ghost
contributions.

The second order form of the Einstein-Hilbert action, when
supplemented by a LM field to impose the classical equations of
motion, also leads to an effective Lagrangian when it is quantized
using the Faddeev-Popov procedure that is invariant under nilpotent
BRST transformation \cite{21,21a}. We expect that as a result, this
theory is restricted to one-loop order in perturbation theory, and is
consistent with unitarity as well as renormalizability.

\end{document}